%% file: drholong.tex
\documentstyle[12pt,epsfig]{article}

\newcommand{\lsim}{\raisebox{-0.13cm}{~\shortstack{$<$ \\[-0.07cm] $\sim$}}~}

\newcommand{\mst}{m_{\tilde{t}}}

\newcommand{\mste}{m_{\tilde{t}_1}}
\newcommand{\mstz}{m_{\tilde{t}_2}}
\newcommand{\msti}{m_{\tilde{t}_i}}

\newcommand{\msb}{m_{\tilde{b}}}

\newcommand{\msbl}{m_{\tilde{b}_L}}
\newcommand{\msbe}{m_{\tilde{b}_1}}

\newcommand{\MstL}{M_{\tilde{t}_L}}
\newcommand{\MstR}{M_{\tilde{t}_R}}
\newcommand{\MsbL}{M_{\tilde{b}_L}}

\newcommand{\msq}{m_{\tilde{q}}}
\newcommand{\msqi}{m_{\tilde{q}_i}}

\newcommand{\SU}{{\mathrm SUSY}}
\newcommand{\oas}{{\cal O}(\alpha_s)}

\newcommand{\dr}{\Delta \rho}
\newcommand{\mt}{m_{t}}
\newcommand{\mb}{m_{b}}
\newcommand{\Mgl}{m_{\tilde{g}}}
\newcommand{\mgl}{m_{\tilde{g}}}
\newcommand{\sq}{\tilde{q}}
\newcommand{\Stop}{\tilde{t}}
\newcommand{\Sbot}{\tilde{b}}
\newcommand{\tst}{\theta_{\tilde{t}}}
\newcommand{\sw}{s_W}
\newcommand{\cw}{c_W}
\newcommand{\stt}{s_{\tilde{t}}}
\newcommand{\ctt}{c_{\tilde{t}}}

\newcommand{\KL}{\left(}
\newcommand{\KR}{\right)}

\newcommand{\Tb}{\tan \beta\hspace{1mm}}

\newcommand{\CTb}{\cot \beta\hspace{1mm}}

\newcommand{\CZb}{\cos 2\beta\hspace{1mm}}

\newcommand{\BE}{\begin{equation}}
\newcommand{\EE}{\end{equation}}
\newcommand{\BEA}{\begin{eqnarray}}
\newcommand{\BEAnn}{\begin{eqnarray*}}
\newcommand{\EEA}{\end{eqnarray}}
\newcommand{\EEAnn}{\end{eqnarray*}}

\def\beq{\begin{equation}}
\def\eeq{\end{equation}}

\def\beqn{\begin{eqnarray}}
\def\eeqn{\end{eqnarray}}
\relax

\def\tb{\tan\beta}

\newcommand{\ra}{\rightarrow}

\newcommand{\s}{\\ \vspace*{-3mm} }

\newcommand{\non}{\nonumber}
\newcommand{\mg}{ m_{\tilde{g}} }

\def\de{\delta}

\def\De{\Delta}

\def\draftdate{\relax}
\def\mda{\relax}
\def\mua{\relax}
\def\mla{\relax}
\def\draft{
\def\thtystars{******************************}
\def\sixtystars{\thtystars\thtystars}
\typeout{}
\typeout{\sixtystars**}
\typeout{* Draft mode!
         For final version remove \protect\draft\space in source file *}
\typeout{\sixtystars**}
\typeout{}
\def\draftdate{\today}
\def\mua{\marginpar[\boldmath\hfil$\uparrow$]%
                   {\boldmath$\uparrow$\hfil}%
                    \typeout{marginpar: $\uparrow$}\ignorespaces}
\def\mda{\marginpar[\boldmath\hfil$\downarrow$]%
                   {\boldmath$\downarrow$\hfil}%
                    \typeout{marginpar: $\downarrow$}\ignorespaces}
\def\mla{\marginpar[\boldmath\hfil$\rightarrow$]%
                   {\boldmath$\leftarrow $\hfil}%
                    \typeout{marginpar: $\leftrightarrow$}\ignorespaces}
\def\Mua{\marginpar[\boldmath\hfil$\Uparrow$]%
                   {\boldmath$\Uparrow$\hfil}%
                    \typeout{marginpar: $\Uparrow$}\ignorespaces}
\def\Mda{\marginpar[\boldmath\hfil$\Downarrow$]%
                   {\boldmath$\Downarrow$\hfil}%
                    \typeout{marginpar: $\Downarrow$}\ignorespaces}
\def\Mla{\marginpar[\boldmath\hfil$\Rightarrow$]%
                   {\boldmath$\Leftarrow $\hfil}%
                    \typeout{marginpar: $\Leftrightarrow$}\ignorespaces}
\overfullrule 5pt
\oddsidemargin -15mm
\marginparwidth 29mm
}

\begin{document}

\def\thefootnote{\fnsymbol{footnote}}

\begin{flushright}
KA--TP--8--1997\\
MPI-PhT/97-39 \\
PM 97/34 \\
hep-ph/9710438 \\
\date{\today}
\end{flushright}

\vspace{1cm}

\begin{center}

{\large\sc {\bf Leading QCD Corrections to Scalar Quark}}

\vspace*{0.4cm} 

{\large\sc {\bf Contributions to Electroweak Precision Observables}}

\vspace{1cm}

{\sc A.~Djouadi$^{1}$, P.~Gambino$^{2}$%
\footnote{Present address: Physik Dept., Technische Universit\"at
M\"unchen, D--85748 Garching.}
, S.~Heinemeyer$^3$, }

\vspace*{.4cm}

{\sc W.~Hollik$^3$, C.~J\"unger$^3$}%
\footnote{Supported by the Deutsche Forschungsgemeinschaft.} 
and {\sc G.~Weiglein$^3$} 

\vspace*{1cm}

$^1$ Physique Math\'ematique et Th\'eorique, UPRES--A 5032, \\
Universit\'e de Montpellier II, F--34095 Montpellier Cedex 5, France. 

\vspace*{0.4cm}

$^2$ Max--Planck Institut f\"ur Physik, Werner Heisenberg Institut, \\
D--80805 Munich, Germany.

\vspace*{0.4cm}

$^3$ Institut f\"ur Theoretische Physik, Universit\"at Karlsruhe, \\
D--76128 Karlsruhe, Germany \\

\end{center}

\vspace*{1cm}

\begin{abstract}

In the supersymmetric extension of the Standard Model we derive the
two--loop QCD corrections to the scalar quark contributions to the
electroweak precision observables entering via the $\rho$ parameter.
A very compact expression is derived for the gluon--exchange contribution. 
The complete analytic result for the gluino--exchange contribution is very 
lengthy; we give expressions for several limiting cases that were derived 
from the general result. The two--loop corrections,
generally of the order of 10 to 30\% of the one-loop contributions, can
be very significant.
Contrary to the Standard Model case, where the QCD 
corrections are negative and screen the one--loop value, the corresponding 
corrections in the supersymmetric case are in general positive, therefore  
increasing the sensitivity in the search for scalar quarks through their 
virtual effects in high--precision electroweak observables. 
 
\end{abstract}

\def\thefootnote{\arabic{footnote}}
\setcounter{footnote}{0}

\newpage

\subsection*{1. Introduction}

Supersymmetric theories (SUSY) \cite{R1} are widely considered as the
theoretically most appealing extension of the Standard Model (SM). They are
consistent with the approximate unification of the three gauge coupling
constants at the GUT scale and provide a way to cancel the quadratic
divergences in the Higgs sector hence stabilizing the huge hierarchy between
the GUT and the Fermi scales. Furthermore, in SUSY theories the breaking
of the electroweak symmetry is naturally induced at the Fermi scale,
and the lightest supersymmetric particle can be neutral, weakly interacting
and absolutely stable, providing therefore a natural solution for the
Dark Matter problem; for recent reviews see for instance Ref.~\cite{R2}. \s

Supersymmetry predicts the existence of scalar partners $\tilde{f}_L, 
\tilde{f}_R$ to each SM chiral fermion, and spin--1/2 partners to the 
gauge bosons and to the scalar Higgs bosons. So far, the direct search of 
SUSY particles at present colliders has not been successful. 
One can only set lower bounds of ${\cal O}(100)$ GeV on 
their masses~\cite{R3}. The search for SUSY particles can be extended to 
slightly larger values in the next runs at LEP2 and at the upgraded 
Tevatron.  To sweep the entire mass range for the SUSY particles, which
from naturalness arguments is expected not to be larger than
the TeV scale, the higher energy hadron or $e^+e^-$ colliders of the next 
decade will be required. \s

An alternative way to probe SUSY is to search for the virtual effects of the 
additional particles. Indeed, now that the top--quark mass is known~\cite{R4},
and its  measured value is in remarkable agreement with the one  
indirectly obtained from high--precision electroweak data,
one can use the available data to search 
for the quantum effects of the SUSY particles: sleptons, squarks, gluinos 
and charginos/neutralinos. In the Minimal Supersymmetric Standard Model (MSSM) 
there are three main possibilities for the virtual effects of SUSY particles 
to be large enough to be detected in present experiments: 

\begin{itemize} 

\item[i)] In the rare decay $b \ra s \gamma$, besides the SM top/W--boson loop
contribution, one has additional contributions from chargino/stop and 
charged Higgs/stop loops \cite{R5}. These contributions can be sizable but 
the two new contributions can interfere destructively in large areas 
of the MSSM parameter space, leading in this case
to a small correction to the decay rate predicted by the SM. 

\item[ii)] If charginos and scalar top quarks are light enough, they can 
affect the partial decay width of the Z~boson into $b$~quarks in a sizable 
way \cite{R6a}. This feature has been widely discussed in the recent years, 
in view of the deviation of the $Z \ra b\bar{b}$ partial width from the 
SM prediction \cite{R6b}. However, for chargino and stop masses beyond 
the LEP2 or Tevatron reach, these effects become too small to be 
observable~\cite{R6b}. 

\item[iii)] A third possibility is the contribution of the scalar quark loops, 
in particular stop and sbottom loops, to the electroweak gauge--boson 
self--energies~\cite{R7a,R7b}: if there is a large splitting between the masses
of these particles, the contribution will grow with the square of the mass of 
the heaviest scalar quark and can be very large. This is similar to the SM case 
where the top/bottom weak isodoublet generates a quantum correction that grows 
as the top--quark mass squared. 

\end{itemize}

In this paper, we will focus on the third possibility and discuss in detail 
the leading contribution 
of scalar quark loops to electroweak precision observables, which  is 
parameterized by their contribution to the $\rho$ parameter. 
The radiative corrections affecting the vector
boson self--energies stemming from charginos, neutralinos and Higgs bosons 
have been discussed in several papers \cite{R7a,R7b}. In the MSSM, because of 
the strong constraints on the Higgs sector, the propagator corrections due to 
Higgs particles are very close to those of the SM for a light Higgs
boson~\cite{hollik}. In the decoupling regime where all scalar Higgs bosons
but the lightest are very heavy, the SUSY Higgs sector is effectively
equivalent to the SM Higgs sector with a Higgs--boson mass of the order of
100 GeV. The contribution of charginos and neutralinos, except from threshold
effects, is also very small \cite{R7b}. 
The main reason is that the custodial symmetry which guarantees that $\rho=1$
at the tree level is only weakly broken in this sector since the terms 
which can break this symmetry in the chargino/neutralino mass matrices 
are all proportional to $M_W$ and hence bounded in magnitude \cite{R7b}. \s

The propagator corrections from squark loops to the electroweak observables 
can be attributed, to a large extent, to the correction to the $\rho$ 
parameter~\cite{R9}, which measures the relative strength of the neutral to 
charged current processes at zero momentum--transfer. This is similar to the 
SM, where the top/bottom contribution to the precision observables is, to a
very good approximation, proportional to their contribution
to the deviation of the $\rho$ parameter from unity. 
Further contributions, compared to the previous one, are suppressed by 
powers of the heavy masses.
It is mainly from this contribution that the top--quark mass has been
successfully predicted from the measurement of the Z--boson observables
and of the W--boson mass at hadron colliders. However, in order for the
predicted value to agree with the experimental one, higher--order radiative
corrections \cite{R10a,R10b,R10c} had to be included. For instance, the 
two--loop 
QCD corrections lead to a decrease of the one--loop result by approximately
$10\%$ and shift the top--quark mass upwards by an amount of $\sim 10$ GeV. 
\s

In order to treat the SUSY loop contributions to the electroweak
observables at the same level of accuracy as the standard contribution,
higher--order corrections should be incorporated. In particular the QCD 
corrections, which because of the large value of the strong coupling 
constant can be rather important, must be known. In a short 
letter~\cite{R11} we have recently presented the results for the
${\cal O}(\alpha_s)$ correction to the contribution of the scalar top
and bottom quark loops to the $\rho$ parameter. In this article we
give the main details of the calculation and present the explicit result
for the gluon--exchange contribution as well as the result for the
gluino--exchange contribution in several limiting cases.\s

The paper is organized as follows. In the next section, we summarize the 
one--loop results and fix the notation. The main features of the 
two--loop calculation are discussed in section~3. In section 4 
a compact expression is given for the gluon--exchange contributions.
The results for the gluino--exchange contributions are presented for
the limiting cases of zero gluino mass and a very heavy gluino as well
as for the case of arbitrary gluino mass but vanishing squark mixing.
Effects of ${\cal O}(\alpha_s)$ corrections to relations between the
squark masses existing in different scenarios are discussed.
In section 5 we give our conclusions.

\subsection*{2. One-loop results}

For the sake of completeness, we summarize in this section the one--loop 
contribution of a squark doublet to the electroweak precision observables. 
Before that, to set the notation, we first discuss the masses and couplings 
of scalar quarks in the MSSM. \s

As mentioned previously, SUSY associates a left-- and a right--handed
scalar partner to each SM quark. The current eigenstates, $\sq_L$ and
$\sq_R$, mix to give the mass eigenstates $\sq_1$ and $\sq_2$; the
mixing angle is proportional to the quark mass and is therefore important
only in the case of the third generation squarks. In the MSSM, the squark
masses are given in terms of 
the Higgs--higgsino mass
parameter $\mu$, the ratio of the vacuum expectation values $\tb$ of the
two Higgs doublet MSSM fields needed to break the electroweak symmetry,
the left-- and right--handed scalar masses $M_{\sq_L}$ and $M_{\sq_R}$
and the soft--SUSY breaking trilinear coupling $A_q$. The top and bottom
squark mass eigenstates and their mixing angles  are determined by
diagonalizing the following mass matrices 
\begin{equation}
{\cal M}^2_{\tilde{t}} = 
\left( 
  \begin{array}{cc} M_{\tilde{t}_L}^2 + m_t^2 + \cos 2 \beta (\frac{1}{2}
                       - \frac{2}{3}s_W^2) \, M_Z^2  & m_t \, M^{LR}_t \\
                    m_t \, M^{LR}_t & M_{\tilde{t}_R}^2 + m_t^2
                                   + \frac{2}{3}\cos 2 \beta \; s_W^2 \, M_Z^2
  \end{array}
\right)
\label{eq:mm1}
\end{equation}
\begin{equation}
{\cal M}^2_{\tilde{b}} = 
\left( 
  \begin{array}{cc} M_{\tilde{b}_L}^2 + m_b^2 + \cos 2 \beta (-\frac{1}{2}
                       +\frac{1}{3}s_W^2) \, M_Z^2  & m_b \, M^{LR}_b \\
                    m_b \, M^{LR}_b & M_{\tilde{b}_R}^2 + m_b^2
                    - \frac{1}{3}\cos 2 \beta \; s_W^2 \, M_Z^2 
  \end{array}
\right) ,
\label{eq:mm2}
\end{equation}
with $M^{LR}_t = A_t - \mu \, \CTb$ and $M^{LR}_b = A_b - \mu \Tb$;
$\sw^2=1-\cw^2 \equiv \sin^2 \theta_W$. Furthermore, SU(2) gauge invariance
requires $M_{\tilde{t}_L} = M_{\tilde{b}_L}$ at the tree--level.
Expressed in terms of the squark masses $m_{\sq_1}, m_{\sq_2}$
and the mixing angle $\theta_{\sq}$ the squark mass matrices read
\begin{equation}
{\cal M}^2_{\sq} = 
\left( 
  \begin{array}{cc} \cos^2 \theta_{\sq} m^2_{\sq_1} + 
                    \sin^2 \theta_{\sq} m^2_{\sq_2} &
                    \sin \theta_{\sq} \cos \theta_{\sq} 
                    (m^2_{\sq_1} - m^2_{\sq_2}) \\
                    \sin \theta_{\sq} \cos \theta_{\sq} 
                    (m^2_{\sq_1} - m^2_{\sq_2}) &
                    \sin^2 \theta_{\sq} m^2_{\sq_1} +
                    \cos^2 \theta_{\sq} m^2_{\sq_2}
  \end{array}
\right) .
\label{eq:squarkmix2}
\end{equation}

Due to the large value of the top--quark mass $m_t$, the mixing between the 
left-- and right--handed top squarks $\tilde{t}_L$ and $\tilde{t}_R$ can be
very large, and after diagonalization of the mass matrix the lightest
scalar top--quark mass eigenstate $\tilde{t}_1$ can be much lighter than the top
quark and all the scalar partners of the light quarks \cite{R12}. The mixing 
in the sbottom sector is in general rather small, except if $A_b, \mu$ 
or $\tb$ are extremely large. In most of our discussion we will assume that,
because of the small bottom mass, the mixing in the sbottom sector is
negligible and therefore $\tilde{b}_L 
\equiv \tilde{b}_1$. \s

Using the notation of the first generation, the contribution of a squark 
doublet $\tilde{u}, \tilde{d}$ to the self--energy of a vector boson
$V\equiv \gamma, Z, W$ and to the $Z$--$\gamma$ mixing is given by the
diagrams of Fig.~\ref{oneloopdiagrams}.  
Summing over all possible flavors and helicities, the squark contribution 
to the transverse parts of the gauge--boson self--energies at arbitrary
momentum--transfer $q^2$ can be written, 
in terms of the Fermi constant $G_F$, as follows:
\begin{eqnarray}
\Pi_{WW}(q^2) &=& - \frac{3 G_F M_W^2}{8 \sqrt{2} \pi^2} \sum_{i,j=1,2}
g_{W \tilde{u}_i \tilde{d}_j }^2 \, \Pi_0 (q^2,
m_{\tilde{u}_i}^2, m_{\tilde{d}_j}^2)
\non \\
\Pi_{ZZ}(q^2) &=& - \frac{3 G_F M_Z^2}{4 \sqrt{2} \pi^2}  \sum_{
\sq= \tilde{u}, \tilde{d}  \atop i,j=1,2} g_{Z \sq_i 
\sq_j }^2 \, \Pi_0 (q^2,m_{\sq_i}^2, m_{\sq_j}^2)
\non \\
\Pi_{Z\gamma}(q^2) &=& - \frac{3 G_F M_Z^2 s_W c_W }{4 \sqrt{2} \pi^2} 
  \sum_{
\sq= \tilde{u}, \tilde{d} \atop i=1,2} g_{Z \sq_i \sq_j } 
e_{\sq_i} \, \Pi_0 (q^2,m_{\sq_i}^2, m_{\sq_i}^2)
\non \\
\Pi_{\gamma \gamma}(q^2) &=& - \frac{3 G_F M_Z^2 s_W^2 c_W^2}
{4 \sqrt{2} \pi^2}   \sum_{
\sq= \tilde{u}, \tilde{d} \atop i=1,2} e_{\sq_i}^2  \, \Pi_0
(q^2, m_{\sq_i}^2, m_{\sq_i}^2),
\label{oneloopgbse}
\end{eqnarray}  
with the reduced couplings of the squarks to the $W$ and $Z$ bosons, including 
mixing $\theta_{\sq}$ between left-- and right--handed squarks, given by
($e_q$ and $I_3^q$ are the electric charge and the weak isospin of the 
partner quark) 
\begin{eqnarray}
g_{W \tilde{u}_i \tilde{d}_j} = 
\left( 
  \begin{array}{cc} \cos \theta_{\tilde u} \cos \theta_{\tilde d} &
- \cos \theta_{\tilde u} \sin \theta_{\tilde d}  \\ 
- \cos \theta_{\tilde u} \sin \theta_{\tilde d} & 
\sin \theta_{\tilde u} \sin \theta_{\tilde d} 
\end{array} \right), \nonumber
\end{eqnarray}
\begin{eqnarray}
g_{Z \sq_i \sq_j} =
\left( 
\begin{array}{cc} (I^q_3-e_q s_W^2) \cos^2\theta_{\sq} - 
e_q s_W^2 \sin^2\theta_{\sq}
&  -I_3^q \sin\theta_{\sq} \cos\theta_{\sq} \\
   -I_3^q \sin\theta_{\sq} \cos\theta_{\sq} & 
-e_q s_W^2 \cos^2\theta_{\sq} + 
(I^q_3-e_q s_W^2) \sin^2\theta_{\sq}   
\end{array} \right) .
\end{eqnarray}

\bigskip
In both the dimensional regularization \cite{R13a} and dimensional reduction 
\cite{R13b} schemes,%
\footnote{In general the dimensional reduction scheme, which preserves
SUSY, should be used. For all quantities considered in this paper
dimensional reduction yields precisely the same result as dimensional
regularization (see however the discussion in section 4.4).}
the function $\Pi_0 (q^2,m_a^2, m_b^2)$ reads
\begin{eqnarray}
\Pi_0 (q^2,m_a^2, m_b^2) = 
\frac4{3} \left[ m_a^2 + m_b^2 - \frac{q^2}{3} +
 \left( m_a^2 + m_b^2 - \frac{q^2}{2} -\frac{(m_a^2-m^2_b)^2}{2\,q^2}
\right)B_0(q^2,m_a,m_b)\right.\non\\
\left.+ \frac{m_a^2-m_b^2}{2\,q^2} (A_0(m_a)-A_0(m_b)) - A_0(m_a)-A_0(m_b)
\right] .
\end{eqnarray}
The Passarino--Veltman one-- and two--point functions \cite{PV} 
are defined as 
\begin{eqnarray}
A_0 (m) &=& m^2 \left[\frac{1}{\epsilon} + 1-\ln \frac{m^2}{\mu^2} +
\epsilon\left(1+\frac{\pi^2}{12}  -\ln \frac{m^2}{\mu^2}
+\frac1{2}\,\ln^2 \frac{m^2}{\mu^2}\right)
\right] \nonumber \\
B_0(q^2,m_a,m_b) &=& \frac{1}{\epsilon} + B_{0}^{\rm fin}(q^2,m_a,m_b)+
\epsilon \,B_{0}^{\epsilon}(q^2,m_a,m_b)\nonumber\\
B_{0}^{\rm fin}(q^2,m_a,m_b)&=&
 2 - \ln \frac{m_a m_b}{\mu^2}
   +\frac{m_a^2-m_b^2}{q^2} \ln\frac{m_a}{m_b} \nonumber \\
& & +\frac{\beta^{1/2}(q^2, m_a^2, m_b^2)}{q^2}
     \ln\frac{m_a^2 + m_b^2 - q^2+ \beta ^{1/2}(q^2, m_a^2, m_b^2)}
{2m_a m_b} , 
\label{eq:A0B0}
\end{eqnarray}
where $\mu$ is the  renormalization scale, $\beta$ the phase space 
function, 
\begin{equation}
\beta(q^2, m_a^2, m_b^2)= q^2-m_a^2-m_b^2+\frac{(m_a^2-m_b^2)^2}{q^2},
\end{equation}
and $2\epsilon =4-n$ with $n$ the space--time dimension. We have 
absorbed a factor $(e^\gamma/4\pi)^\epsilon$, with $\gamma$ the Euler constant,
in the 't Hooft scale $\mu$
to prevent uninteresting combinations of $\ln 4\pi,\ \gamma \ldots$ in 
our results.
The explicit form of the function $B_{0}^{\epsilon}$ is not 
needed for our purposes but can be found in Ref.~\cite{B0eps}. \s

At zero momentum--transfer, $q^2=0$, the function $\Pi_0$ is finite and
reduces to
\beq
 F_0(m_a^2, m_b^2) \equiv
\Pi_0(0,m_a^2, m_b^2) =
m_a^2+m_b^2  - \frac{2m_a^2 m_b^2 } {m_a^2- m_b^2 } \ln \frac{m_a^2}
{m_b^2} .
\label{oneloopresult}
\eeq

We now focus on the contribution of a squark doublet to the $\rho$
parameter. It is well-known that the deviation of the $\rho$ parameter
from unity parameterizes the leading universal corrections induced by
heavy fields in electroweak amplitudes. It is due to a mass splitting
between the fields in an isospin doublet. Compared to this correction 
all additional  contributions are suppressed.
In the relevant cases of the W--boson mass and of the effective weak
mixing angle $\sin^2 \theta_W^{\rm eff}$, for example, a doublet of heavy
squarks would induce shifts proportional to its contribution to $\rho$,
\beq
\delta M_W \approx \frac{M_W}{2} \frac{\cw^2}{\cw^2-\sw^2} \Delta\rho;
\ \ \ \ 
\delta\sin^2\theta_W^{\rm eff}\approx - \frac{\cw^2\sw^2}{\cw^2-\sw^2} \Delta\rho.
\label{shifts}
\eeq

In terms of the transverse parts of the W-- and Z--boson self--energies
at zero momentum--transfer, the squark loop contribution to the $\rho$
parameter is given by
\beq
\rho = \frac{1}{1-\Delta \rho} \ ; \ \ \ \ \Delta \rho =
\frac{\Pi_{ZZ}(0)}{M_Z^2} - \frac{\Pi_{WW}(0)}{M_W^2}  \ .
\eeq
Using the previous expressions for the W-- and Z--boson self-energies 
and neglecting the mixing in the sbottom sector, one obtains for the 
contribution of the $\tilde{t}/\tilde{b}$ doublet at one--loop order
(only the left--handed sbottom $\tilde{b}_L$ contributes for
$\theta_{\tilde{b}}=0$) :
\begin{eqnarray}
\Delta \rho_0^\SU  &=& \frac{3 G_F}{8 \sqrt{2} \pi^2} 
\left[ -\sin ^2 \theta_{\tilde{t}}
\cos^2 \theta_{\tilde{t}} \ F_0( m_{\tilde{t}_1}^2,   m_{\tilde{t}_2}^2) 
+ \cos^2 \theta_{\tilde{t}} \ F_0( m_{\tilde{t}_1}^2,   m_{\tilde{b}_L}^2) 
\right.
\non \\
&& \left. {} + 
\sin^2 \theta_{\tilde{t}} \ F_0( m_{\tilde{t}_2}^2,   m_{\tilde{b}_L}^2) 
\right] .
\label{drhooneloop}
\end{eqnarray}
The function $F_0$ vanishes when the two squarks running in the loop are 
degenerate in mass, $F_0(m_q^2, m_q^2)=0$. In the limit of large squark mass
splitting it becomes proportional to the heavy squark mass squared:
$F_0(m_a^2,0)=m_a^2$. Therefore, the contribution of a squark doublet
becomes in principle very large when the mass splitting between squarks is
large. This is exactly the same situation as in the case of the SM where
the top/bottom contribution to the $\rho$ parameter at one--loop order,
keeping both the $t$ and $b$ quark masses, reads~\cite{R9}
\begin{eqnarray}
\Delta \rho_0^{\rm SM}  = \frac{3 G_F}{8 \sqrt{2} \pi^2}
F_0(m_t^2, m_b^2).
\end{eqnarray}
For $\mt \gg m_b$ this  leads to the well--known quadratic correction
$\Delta \rho_0^{\rm SM}= 3 G_F m_t^2 /(8\sqrt{2}\pi^2)$. \s

In Fig.~\ref{oneloopdrhoa} we display the one--loop correction to the $\rho$
parameter that is induced by the $\tilde{t}/\tilde{b}$ isodoublet. 
The scalar mass parameters are assumed
to be equal, $M_{\tilde{t}_{L}} = M_{\tilde{t}_{R}}$ and
$M_{\tilde{b}_{L}} = M_{\tilde{b}_{R}}$, as it is approximately
the case in Supergravity models with scalar mass unification at the GUT
scale \cite{SUGRA}. As mentioned above, SU(2) gauge invariance 
requires at the tree--level%
\footnote{The corrections to 
the relations between
the squark masses will be discussed in subsection 4.4.}
$M_{\tilde{t}_{L}} = M_{\tilde{b}_{L}}$, yielding in this
case $M_{\tilde{t}_{L}}=M_{\tilde{t}_{R}}=M_{\tilde{b}_{L}} = M_{\tilde{b}_{R}}
= m_{\sq}$. In this scenario, the scalar top mixing angle is either
very small, $\theta_{\tilde t} \sim 0$, or almost maximal,
$ \theta_{\tilde t} \sim  -\pi/4$, in most of the MSSM parameter space.
The contribution $\Delta \rho_0^\SU$ is shown as a function of the common
squark mass $m_{\sq}$ for $\tb=1.6$ for the two cases 
$M^{LR}_t=0$ (no mixing) and $M_t^{LR}=200$ GeV (maximal mixing);%
\footnote{As will be discussed below, the case of exact maximal mixing,
$ \theta_{\tilde t} = -\pi/4$, is not possible in the scenario
with $M_{\tilde{t}_{L}}=M_{\tilde{t}_{R}}$ and $\Tb = 1.6$. Since
$\theta_{\tilde t}$ is already very close to the maximal value for
$M^{LR}_t = 200$~GeV, we will in the following refer to this 
scenario as ``maximal mixing''.} 
the bottom mass and therefore the mixing in the sbottom sector are
neglected leading to $\msbl = \msbe \simeq \msq$.
Here and in all the numerical calculations in this paper we use 
$m_t=175$~GeV, $M_Z=91.187$~GeV,  $M_W=80.33$~GeV, and $\alpha_s=0.12$. 
The electroweak mixing angle is defined from the ratio of the vector
boson masses: $\sw^2= 1- M_W^2/M_Z^2$. \s

As can be seen, the  correction is rather large for small $\msq$,
exceeding the level $\dr = 1.3 \times 10^{-3}$ of experimental 
sensitivity\footnote{The correction to $\Delta\rho$ discussed here directly
corresponds to a correction to the effective parameter $T$ defined in 
Ref.~\cite{PT}, or equivalently to $\epsilon_1$ as given in Ref.~\cite{AB} 
or the combination of parameters defined in Ref.~\cite{Schild}.
Using the 1997 precision data, the resolution on  $\epsilon_1$ is estimated
to be $1.3 \times 10^{-3}$ \cite{AB2}. A similar estimate can be readily 
obtained using the present experimental errors on the world average
\cite{EXP} of $M_W$, 80 MeV,  and $\sin^2\theta_W^{\rm eff}$, 
$2.2\times 10^{-4}$, in eq.~(\ref{shifts}).}
in the case of no mixing for $m_{\sq} \sim 150$ GeV, which corresponds to
the experimental lower bound on the common squark mass \cite{R3},
and getting very close to it in the case of maximal mixing. 
For large $m_{\sq}$ values, the two stop and the sbottom masses
are approximately degenerate since  $m_{\sq} \gg m_t$, and the 
contribution to $\Delta \rho$ becomes very small. \s

For illustration, we have chosen the value $\Tb=1.6$ which is favored
by $b$--$\tau$ Yukawa coupling unification scenarios \cite{SUGRA}. In fact, 
the analysis 
depends only marginally on $\tb$ if $M^{LR}_t$ (and not $A_t$ and $\mu$) is 
used as input parameter. The only effect of varying $\Tb$ is  then to 
slightly alter the D--terms in the mass matrices eqs.~(1--2), which does not
change the situation in a significant way. However, for large $\tb$
values, $\Tb \sim m_t/m_b$, the mixing in the sbottom sector has to be
taken into account, rendering the analysis somewhat more involved. 
We will focus on the scenario with a low value of $\Tb$ in the
following. \s
 
In Fig.~\ref{oneloopdrhob} we display $\dr_0^\SU$ as a function of the
mixing angle for three values of the common squark mass, $m_{\sq} =150$,
250 and 500~GeV, and for $\Tb=1.6$.  The contribution is practically flat
except for values of the mixing angle very close to the lower limit, 
$\theta_{\tilde{q}} \sim -\pi/4$. This justifies the choice of concentrating 
on the two extreme cases. In fact, in the case $M_{\tilde{t}_L}=
M_{\tilde{t}_R}$, the maximal mixing scenario $\theta_{\tilde{q}} = -\pi/4$ 
is only obtained exactly
when the D--terms are set to zero, which is the case when $\tb=1$. 
One can also have maximal mixing if the sums of 
$M_{\tilde{t}_L}^2$ and $M_{\tilde{t}_R}^2$
with their corresponding D--terms are equal, 
making the diagonal entries in the mass matrices eqs.~(1--2) 
identical. \s

The parameter $M_{t}^{LR}$ does not only influence the mixing but
also the mass splitting between the scalar top quarks. The effect of
varying $M_{t}^{LR}$ on $\Delta \rho_0^{\rm SUSY}$ is displayed in 
Table~1 for the case of exact maximal mixing, i.e.\ $\Tb = 1$.
The scenario with $\Tb = 1.6$ yields similar numerical results.
For large values of $M_{t}^{LR}$ the contribution to the 
$\rho$ parameter can become huge, exceeding by far the level of
experimental observability. The increase of $\Delta \rho_0^{\rm SUSY}$
with larger values of $M_{t}^{LR}$ is due to the increased mass
splitting between the two stop masses and $\msbe$ 
that is induced for $m_t M_{t}^{LR} \gg m_{\tilde{q}}^2$. 
For values of $m_{\tilde{q}}$ 
comparable to the top quark mass, $m_{\tilde{q}} \lsim 200$~GeV, 
the mass splitting is already large even for small $M_{t}^{LR}$ due 
to the additional $m_t^2$ term in the stop mass matrix; this is similar 
to the no--mixing case. \s

If the GUT relation $M_{\tilde{t}_L} \simeq M_{\tilde{t}_R}$ is relaxed and
large values of the mixing parameter $M_t^{LR}$ are assumed, the splitting
between the stop and sbottom masses is so large that the contribution to the
$\rho$ parameter can become even bigger than in the previously
discussed scenario (see subsection 4.4 for a more detailed discussion).
\s

\begin{table}[hbt]
\renewcommand{\arraystretch}{1.5}
\begin{center}
\begin{tabular}{|c||c|c|} \hline
$m_{\tilde{q}}$ (GeV) & $M_t^{LR}$ (GeV) & $\Delta \rho_0^{\rm SUSY} 
\times 10^{-3}$ \\ \hline \hline
200 & 100 & 1.68 \\ 
    & 200 & 1.35 \\
    & 300 & 0.89 \\
    & 400 & 1.28 \\ \hline
500 & 200 & 0.34 \\
    & 500 & 0.23 \\
    & 1400& 2.04 \\
    & 1550& 4.97 \\ \hline 
800 & 500 & 0.12 \\
    & 1500& 0.07 \\
    & 2000& 0.30 \\
    & 3700& 15.8  \\ \hline  
\end{tabular}
\renewcommand{\arraystretch}{1.2}
\caption[]{\small $\Delta \rho_0^{\rm SUSY}$ in units of $10^{-3}$
for several values of $m_{\tilde{q}}$ and $M_t^{LR}$ (in GeV) and
$\Tb = 1$. The 
values of $M_t^{LR}$ are chosen such that the corresponding squark
masses lie in the experimentally allowed range.}
\end{center}
\end{table}

\subsection*{3. Two--loop calculation}

\subsubsection*{3.1. Renormalization}

The QCD corrections to the squark contributions to the vector
boson self--energies, Fig.~\ref{twoloopdiagrams}, can be divided into
three different classes: the pure scalar diagrams
(Fig.~\ref{twoloopdiagrams}a--c), the gluon exchange diagrams
(Fig.~\ref{twoloopdiagrams}d--j), and the gluino--exchange diagrams
(Fig.~\ref{twoloopdiagrams}k--n). These diagrams have to be supplemented
by counterterms for the squark and quark mass renormalization
(Fig.~\ref{twoloopcts}a--c) as well as for the renormalization
of the squark mixing angle, (Fig.~\ref{twoloopcts}d). The three different
sets of contributions together with the respective counterterms 
are separately gauge--invariant and ultraviolet finite. For the gluon
exchange contribution we have only considered the squark loops, since the
gluon exchange in quark loops is just the SM contribution, yielding the result
$\Delta \rho ^{\rm SM}_1 = - \Delta \rho_0^{\rm SM} \frac{2}{3}
\frac{\alpha_s}{\pi} (1+\pi^2/3 )$~\cite{R10a}. \s

As mentioned above, the results presented in the following are
precisely the same in dimensional regularization as in dimensional
reduction. The renormalization procedure is performed as 
follows. We work in the on--shell scheme where the quark and squark 
masses are defined as the real part of the pole of 
the corresponding propagators. One further needs a
prescription for the renormalization of the squark mixing angle. The
renormalized mixing angle can be defined by requiring that the renormalized
squark mixing self--energy $\Pi_{\sq_1 \sq_2}^{\rm ren}(q^2)$ vanishes at a 
given momentum--transfer $q_0^2$, for example when one of the two squarks is 
on-shell. This means that the two squark mass eigenstates $\tilde{q_1}$ and
$\tilde{q_2}$ do not mix but propagate independently for this value
of $q^2$. Expressing the parameters in eq.~(\ref{eq:squarkmix2})
by renormalized quantities and choosing the field renormalization of the
squarks appropriately, this renormalization condition yields for the mixing
angle counterterm
\begin{equation}
\delta \theta_{\sq} (q_0^2)= \frac{1}{m^2_{\sq_1} - m^2_{\sq_2}}
\Pi_{\sq_1 \sq_2}(q_0^2) .
\end{equation}
Finally, we have also included, as a check, the field renormalization 
constants of the quarks and the squarks in our calculation; they of course 
have to drop out in the final result, which we have verified by explicit 
calculation. \s

The one--loop diagrams  of Fig.~\ref{massct}a--c provide the
renormalization of the squark masses and the squark wave functions.
As the one--loop squark contributions to the vector boson self-energies
are finite at vanishing external momentum (see eq.~(\ref{oneloopresult})),
we notice that the $O(\epsilon)$ part of the squark mass counterterm is 
not needed. We have contributions from the three diagrams of
Fig.~\ref{massct}a, b, and c involving gluon exchange, gluino
exchange and a pure scalar contribution, respectively.
The explicit form of the pure scalar contribution is not needed
as we will see later. The 
contribution of the gluon exchange to the squark mass
counterterm is given by
\begin{equation}
\msqi \delta^{g} \msqi  =  
-\, \frac{\alpha_s}{2 \pi}\,\msqi^2 \left[\frac1{\epsilon} + \frac7{3}
- \ln \frac{\msqi^2}{\mu^2} \right],
\end{equation}
while the one of the gluino exchange reads
\begin{eqnarray}
\msqi \delta^{\tilde{g}} \msqi & = & 
-\,\frac{\alpha_s}{ 3\pi} \left[ (m_q^2 +\mgl^2-\msqi^2) \;
   B_0(\msqi^2,m_q,\mgl) +A_0(\mgl) + A_0(m_q)  \right. \non \\
& & + \left. 2(-1)^i \sin2 \theta_{\tilde{q}}  \;\mg \, m_q \;
B_0(\msqi^2,m_q,\mgl) \right],
\end{eqnarray}
where $A_0$ and $B_0$ have been defined previously, and are needed only up 
to $O(1)$ in the $\epsilon $ expansion. Although in our renormalization
scheme the contribution of the quartic--squark interaction to the squark 
masses drops out in the final result, we will also give its expression 
for later convenience ($i'=3-i$):
\begin{eqnarray}
\msqi \delta^{\tilde{q}} \msqi = \frac{\alpha_s}{ 3\pi} \left[
    \cos^2 2 \theta_{\tilde{q}} A_0(m_{\tilde{q}_i})
+   \sin^2 2 \theta_{\tilde{q}} A_0(m_{\tilde{q}_{i'} }) \right] .
\end{eqnarray}
Concerning the quark  mass counterterm, Fig.~\ref{massct}d,
in principle the $O(\epsilon)$ term is needed because the quark
loop contributions to the vector boson self-energies are ultraviolet divergent
even at $q^2=0$. However, in $\dr$ this contribution drops out 
as the one--loop quark contribution to this physical quantity is finite.
As mentioned above, the gluon contribution to the quark mass counterterm
is only relevant for the pure SM correction and is not needed in the
present context. The gluino contribution can be expressed as
\begin{eqnarray}
 \delta^{\tilde{g}} m_q & = &  \frac{\alpha_s}{3\pi} \sum_{i=1,2}\Big[ 
 (-1)^i  \, \sin 2 \theta_{\tilde q} \; 
      \mgl \; B_0(m_q^2, \mgl, \msqi)\nonumber\\
&& +
\frac1{2\, m_q}\left( (m_q^2 + \mgl^2-\msqi^2) B_0(m_q^2, \mgl, \msqi)
+ A_0(\msqi)- A_0(\mgl)\right)           
 \Big].
\label{quarkct}
\end{eqnarray}
Finally, the counterterm for the squark mixing angle, defined at a given
$q_0^2$, is given by 
\begin{eqnarray}
\delta\theta_{\tilde t} (q_0^2) = \frac{\alpha_s}{3 \pi} 
\frac{\cos 2 \theta_{\tilde t} } {\mste^2 - \mstz^2} 
\left[ 4 \mt \mgl B_0(q_0^2, \mt, \mgl) + \sin 2\theta_{\tilde t} 
(A_0(\mstz) - A_0(\mste) ) \right] \ .
\end{eqnarray}
As discussed above, for the value of $q_0^2$ one can either choose 
$m^2_{\tilde{t}_1}$ or $m^2_{\tilde{t}_2}$, the difference being very small.
In our analysis we have chosen $q_0^2 = m^2_{\tilde{t}_1}$.
This renormalization condition is equivalent to the one used in 
Refs.~\cite{R13} for scalar quark decays.

\bigskip 
 
Let us now discuss the separate contributions of the various diagrams. 
The contribution of the pure scalar diagrams vanishes, while for the 
gluon--exchange diagrams one needs to calculate only the first four
genuine two--loop diagrams, Fig.~\ref{twoloopdiagrams}d--g,
and the  corresponding counterterm with the mass renormalization insertion, 
Fig.\ref{twoloopcts}a. The other diagrams do not contribute for
the following reasons: 

\begin{itemize}
\item[(a)]
The diagram Fig.~\ref{twoloopdiagrams}h is exactly canceled by the
corresponding diagram with the mass counterterm insertion,
Fig.~\ref{twoloopcts}b. It should be noted that the expression for
$B_0^{\epsilon}$ is needed in order to obtain this result. 

\item[(b)] 
The reducible diagram of Fig.~\ref{twoloopdiagrams}a involving the quartic
squark interaction contributes only to the longitudinal components of 
the vector boson self--energies and can therefore be discarded. 

\item[(c)]
The diagrams Fig.~\ref{twoloopdiagrams}b--c are canceled by the
corresponding diagrams with the counterterms, Fig.~\ref{twoloopcts}a--b, 
for mass renormalization (for the diagonal terms) 
and mixing angle renormalization (for the non-diagonal terms). 
The diagrams in Fig.~\ref{twoloopcts}d--f for the vertex corrections
contain only field renormalization constants which drop out in the final
result.

\item[(d)]
The gluon tadpole--like diagrams of Fig.~\ref{twoloopdiagrams}i--j 
give a vanishing contribution in both dimensional regularization and reduction.

\end{itemize}

For the gluino--exchange diagrams, one has to calculate all diagrams
of the types shown in Fig.~\ref{twoloopdiagrams}k--n and their
corresponding counterterm diagrams depicted in Fig.~\ref{twoloopcts}.

\bigskip 

We now briefly describe the evaluation of the two--loop diagrams.
As explained above, we have both irreducible two--loop diagrams at zero
momentum--transfer and counterterm diagrams. After reducing their tensor
structure, they can be decomposed into two--loop scalar integrals at zero 
momentum--transfer (vacuum integrals) and products of one--loop integrals.
The vacuum integrals are known for arbitrary internal masses and
admit a compact representation for $\epsilon \to 0$  in terms of logarithms
and dilogarithms (see for instance Ref.~\cite{Davydychev}), while the 
one--loop
integrals $A_0, B_0$ are well--known (see eq.~(\ref{eq:A0B0})). We have
used two independent implementations of the various steps of this procedure
and obtained identical results. \s

In the first implementation, the diagrams were generated with the Mathematica
package {\em FeynArts} \cite{FA}. The model file contains, besides the SM
propagators and vertices, the relevant part of the MSSM Lagrangian,
i.e.\ all SUSY propagators ($\tilde{t}_1, \tilde{t}_2, \tilde{b}_1, \tilde{b}_2,
\tilde{g}$) needed for the QCD--corrections and the appropriate
vertices (gauge boson--squark vertices, squark--gluon and squark--gluino
vertices). The program inserts propagators and vertices into the 
graphs in all possible ways and creates the amplitudes including all
symmetry factors. The evaluation of the two--loop diagrams and counterterms
was performed  with the Mathematica package {\em TwoCalc}~\cite{TC}.
By means of two--loop tensor integral decompositions it reduces
the amplitudes to a minimal set of standard scalar integrals, consisting
in this case of the basic one--loop functions $A_0, B_0$ (the $B_0$
functions originate from the counterterm contributions only) and the
genuine two--loop function $T_{134}$~\cite{Davydychev}, i.e.\ the
two--loop vacuum integral. As a check of our calculation, the transversality
of the two--loop photon and $\gamma Z$ mixing self--energies at arbitrary
momentum transfer and the vanishing of their transverse parts at $q^2 = 0$
was explicitly verified with {\em TwoCalc}. Inserting the explicit
expressions for $A_0, B_0, T_{134}$ in the result for $\Delta\rho$,
the cancellation of the $1/\epsilon^2$ and $1/\epsilon$ poles
was checked algebraically, and a result in terms of logarithms and
dilogarithms was derived. {}From this output a Fortran code was
created which allows a fast calculation for a given set of
parameters. \s

In the second implementation, completely independent, the diagrams were
not generated automatically, but the analytic simplifications and the
expansions in the limiting cases (small and large gluino mass, maximal
and minimal mixing) were 
carried out by using the Mathematica package
{\it ProcessDiagram}~\cite{ProcD}. In this way the results can be cast
into a relatively compact form, shown in the following.

\subsection*{4. Two--loop results} 

\subsubsection*{4.1. Gluon exchange}

In order to discuss our results, let us first concentrate on the 
contribution of the gluonic corrections and the corresponding counterterms.
At the two--loop level, the results for the electroweak gauge--boson 
self--energies at zero momentum--transfer have very simple 
analytical expressions. In the case of an isodoublet $(\tilde{u}, 
\tilde{d})$ where general mixing is allowed, the structure is similar 
to eq.~(\ref{oneloopgbse}) and eq.~(\ref{oneloopresult}) with the
$g_{V\sq_i\sq_j}$ as given previously: 
\beqn
\Pi_{WW} (0)
& = & - \frac{G_F M_W^2 \alpha_s}{4 \sqrt{2} \pi^3} \, \sum_{i,j=1,2} 
  g_{W\tilde{u}_i\tilde{d}_j}^2  \, F_1 \left( 
m_{\tilde{u}_i}^2, m_{\tilde{d}_j}^2 \right), \nonumber \\
\Pi_{ZZ}(0)
& = &- \frac{G_F M_Z^2 \alpha_s}{2 \sqrt{2} \pi^3}   
\sum_{\sq= \tilde{u},\tilde{d} \atop i,j=1,2} 
g_{Z\sq_i\sq_j}^2  F_1 \left( m_{\sq_i}^2, m_{\sq_j}^2 
\right) .
\end{eqnarray}
The two--loop function $F_1(x,y)$ is given in terms of dilogarithms by
\begin{equation}
F_{1}(x,y) = x+y- 2\frac{xy}{x-y} \ln \frac{x}{y} \left[2+
\frac{x}{y} \ln \frac{x}{y} \right] 
+\frac{(x+y)x^2}{(x-y)^2}\ln^2 \frac{x}{y} 
-2(x-y) {\rm Li}_2 \left(1-\frac{x}{y} \right) . 
\end{equation}
This function is symmetric in the interchange of $x$ and $y$.
As in the case of the one--loop function $F_0$, it vanishes for 
degenerate masses, $F_1(x,x)=0$, while in the case of large 
mass splitting it increases with the heavy scalar quark mass 
squared: $F_1 (x,0) = x( 1 +\pi^2/3)$.  \s

{}From the previous expressions, the contribution of the $(\tilde{t}, 
\tilde{b})$ doublet to the $\rho$ parameter, including the two--loop 
gluon exchange and pure scalar quark diagrams are obtained 
straightforwardly. In the case where the $\tilde{b}$ mixing is neglected, 
the SUSY two--loop contribution is given by an expression similar to 
eq.~(\ref{drhooneloop}):
\beqn
\dr^\SU_{1, \rm gluon} &=& \frac{G_F \alpha_s}{4 \sqrt{2} \pi^3} \left[ 
- \sin^2\theta_{\tilde{t}} \cos^2\theta_{\tilde{t}}  
F_1\left( \mste^2,  \mstz^2 \right) \right. \nonumber \\ 
&& {} \left. + \cos^2 \theta_{\tilde{t}} 
F_1 \left( \mste^2, \msbl^2 \right)
+\sin^2 \theta_{\tilde{t}}  
F_1 \left( \mstz^2, \msbl^2 \right) \right]. 
\eeqn
The two--loop gluonic SUSY contribution to $\dr$ is shown in
Fig.~\ref{twoloopgluonplota} as a function of the common scalar mass
$m_{\sq}$ for the two scenarios discussed previously: $\theta_{\tilde{t}} = 0$
and $\theta_{\tilde{t}} \simeq -\pi/4$. As can be seen, the two--loop
contribution is of the order of 10 to 15\% of the one--loop result. Contrary
to the SM case (and to many QCD corrections to electroweak processes in the SM,
see Ref.~\cite{GK} for a review) where the two--loop correction screens 
the one--loop contribution, $\Delta \rho_{1, \rm gluon}^\SU$ has the same sign 
as $\Delta \rho_0^\SU$. For instance, in the case of degenerate 
scalar top quarks with masses $\mst \gg \msb$, the result is the same as
the QCD correction to the $(t,b)$ contribution in the SM, but with 
opposite sign, see section 4.3 below.
The gluonic correction to the contribution of scalar quarks to the $\rho$
parameter will therefore enhance the sensitivity in the search of the virtual
effects of scalar quarks in high--precision electroweak measurements. 
The dependence of the two--loop gluonic contribution on the stop mixing 
angle $\theta_{\tilde t}$ exhibits the same behavior as the one--loop 
correction: $\dr^\SU_{\rm 1, gluon}$ is nearly constant for
all possible values of $\theta_{\tilde t}$; only in the region of
maximal mixing it decreases rapidly. 

\subsubsection*{4.2. Gluino exchange}

Like for the gluon--exchange contribution we have also derived a
complete analytic result for the gluino--exchange contribution. The
complete result is however very lengthy and we therefore present here
explicit expressions only for the limiting cases of light and heavy
gluino mass, and for the case of no squark mixing. The complete expression
is available in Fortran and Mathematica format from the authors. \s

In order to make our expressions as compact as possible, we  use $s_{\tilde t} 
\equiv \sin \theta_{\tilde t}$ and $c_{\tilde t} \equiv \cos\theta_{\tilde t}$ 
as abbreviations and introduce the following notation,
\beqn
d_{xy} = (m_x^2 - m_y^2) \ , \ \
 \mbox{where} \quad   m_1 = \mste, \; m_2 = \mstz, \; m_L = \msbl, \; 
 m_g = m_{\tilde{g}} . \non
\eeqn
The gluino--contribution for vanishing gluino mass is given by
\input delrhomglnull

The result for $\mgl = 0$ is compared to the complete result of the
gluino contribution for $\mgl = 10$, 200, 500 GeV in
Figs.~\ref{mglzeroplota}, \ref{mglzeroplotb}. In Fig.~\ref{mglzeroplota} 
the results for the no mixing scenario are displayed, while
Fig.~\ref{mglzeroplotb} shows the maximal mixing case. As expected, the
curves for $\mgl = 0$ and $\mgl = 10$ GeV are very close, while 
the curves for large gluino masses significantly deviate from the 
$\mgl = 0$ case. For the no mixing scenario the light gluino expression
of eq.~(\ref{eq:lightgluino}) reproduces the exact result within
5$\times 10^{-5}$ for $\mgl<90,130,350$ GeV in the cases of
$m_{\sq}=100,250,500$ GeV, respectively. In the maximal mixing
scenario the correction is much smaller for a light gluino than in
the no mixing case; the light gluino expression reproduces the exact 
result within 5$\times 10^{-5}$ for $\mgl<20,300,400$ GeV in the cases
of  $m_{\sq}=100,250,500$ GeV and maximal mixing. \s

As second limiting case we give a series expansion in powers of the 
inverse gluino mass up to ${\cal O}(1/\mgl^3)$.  Note that, due to the 
term linear in $\mgl$ in eq.~(\ref{quarkct}), the expansion actually 
starts at ${\cal O}(1/\mgl)$. 
\pagebreak
\input delrhomglseries

Fig.~\ref{mglseriesplota} shows the quality of the heavy gluino
expansion up to ${\cal O}(1/\mgl^3)$ in the two cases of minimal
and maximal mixing for $m_{\sq}=100$ GeV. The expansion of
eq.~(\ref{heavygluino}) reproduces the exact results within at most 30\%
(which corresponds to a maximum deviation of about $5\times 10^{-5}$ in
$\Delta\rho$) for $\mgl>200$ GeV in the no mixing case and for
$\mgl>340$ GeV in the maximal mixing scenario.  
We have also verified that higher orders in the expansion improve
significantly the convergence of the series, but we have not included
unnecessary long expressions. As can be expected, when the value of
$m_{\sq}$ is increased, the quality of the heavy gluino approximation
deteriorates. At the same time, however, the gluino correction becomes
smaller and can mostly be neglected; it never exceeds 1.5 $\times 10^{-4}$
and 1$\times 10^{-4}$ for $m_{\sq}>200$ GeV in the cases of minimal and 
maximal mixing, respectively 
(see Figs.~\ref{mglzeroplota} and \ref{mglzeroplotb}).
We conclude that in the region of the parameter space where the gluino
contribution is relevant, and the gluino is not too
light (say $\mgl>300$), eq.~(\ref{heavygluino}) approximates the full 
expression sufficiently well. \s

As mentioned above, for most of the parameter space the mixing in
the stop sector is either zero or nearly maximal. As a third limiting
case we give $\dr^\SU_{1, {\mathrm gluino}}$ for arbitrary $\mgl$, 
but with zero
stop mixing. The result in this limiting case has already been
displayed in Fig.~\ref{mglzeroplota} and Fig.~\ref{mglseriesplota}
(solid curve).
\input delrhonomix
Here we have used $ \lambda(x,y,z)\equiv  \sqrt{(z- x - y)^2 - 4 \,x \,y}$
and the function $\Phi(x,y,z)$ is defined according to the sign of 
$\lambda(x,y,z)^2$. For $\lambda(x,y,z)^2\ge 0$,
\BEA
\Phi(x,y,z)&=&\frac{z}{\lambda}
  \left\{2 \ln\left(\frac{z+x -y-\lambda}{2\,z}\right) 
                \ln\left(\frac{z+y-x-\lambda}{2\,z}\right) 
                -\ln\frac{x}{z} \ln\frac{y}{z} \right.\nonumber\\ &&
\left.-
        2 {\rm Li_2}\left(\frac{z+x -y-\lambda}{2\,z}\right) -
        2 {\rm Li_2}\left(\frac{z+y-x-\lambda}{2\,z}\right)
        + \frac{\pi^2}{3}\right\},
\EEA
while for $\lambda(x,y,z)^2\le 0$
\BEA
\Phi(x,y,z)& =&  \frac{2\,z}{\sqrt{-\lambda^2}}
\left\{ {\rm Cl}_2\left[
2\, {\rm arccos}\left(\frac{x + y -z }{2\,\sqrt{x\, y}}\right)\right] 
+   {\rm Cl}_2\left[
2\, {\rm arccos}\left(\frac{x - y +z}{2\,\sqrt{x\,z}}\right)\right]
\right.\nonumber\\&& \left.+
        {\rm Cl}_2\left[
2\, {\rm arccos}\left(\frac{z-x + y}{2\,\sqrt{y\,z}}\right)
               \right] \right\},
\EEA
where ${\rm Cl}_2(x)= {\rm Im} \,{\rm Li_2} (e^{ix})$ is the Clausen function;
limiting expressions and additional details on this function can be found
in Ref.~\cite{Davydychev}.

\subsubsection*{4.3. Supersymmetric limit}

The sign of the gluonic two--loop contributions is, as discussed above,
always the same as the sign of the one--loop contribution, contrary to the
case of the two--loop contributions in the SM. In the limit of 
vanishing gluino mass, $\mste = \mstz = \mt$, $\msbl = m_b = 0$
(supersymmetric limit), the gluon
exchange contribution of the scalar quarks reads
\beq
\dr^{\rm SUSY}_{\rm 1, gluon} = \frac{\alpha_s G_F \mt^2}{\sqrt{2} \pi^3}
     \frac{3 + \pi^2}{12} = \dr^{\rm SM}_0 \frac{2}{3}
     \frac{\alpha_s}{\pi} \left(1 + \frac{\pi^2}{3} \right),
\label{susylimit}
\eeq
which exactly cancels the quark loop contribution \cite{R10a}. The gluino
exchange contribution in this limit is given by
\beq
\dr^{\rm SUSY}_{\rm 1, gluino} = - \dr ^{\rm SM}_0 \frac{8}{3}
     \frac{\alpha_s}{\pi},
\eeq
which numerically cancels almost completely the contribution of 
eq.~(\ref{susylimit}).

\subsubsection*{4.4. Corrections to the squark masses}

The analytical formulas for $\Delta \rho$ given in the previous sections
are exclusively expressed in terms of the physical squark masses. The
formulas therefore allow a general analysis since they do not rely on any
specific model assumption for the mass values. For our numerical
analysis, however, since the physical values of the squark masses are
unknown, we had to calculate the masses from the (unphysical) soft--SUSY
breaking mass parameters.
We have concentrated on the MSSM scenario with soft--breaking terms
obeying the SU(2) relation $\MstL = \MsbL$. 
The $\Stop$ and $\Sbot$ masses are obtained by diagonalizing the
tree--level mass matrices. Using eqs.~(\ref{eq:mm1}--\ref{eq:squarkmix2}), 
the masses $m_{\tilde{t}_1}$, $m_{\tilde{t}_2}$ and
$m_{\tilde{b}_L} = m_{\tilde{b}_1}$%
\footnote{Since we set $\mb = 0$, thus neglecting mixing in the
$\Sbot$ sector, we have $\msbe = \msbl$, and $m_{\tilde{b}_2}$
decouples from the $\rho$ parameter.}
as well as the mixing angle $\theta_{\tilde t}$ can be expressed in
terms of the three parameters $\MstL$, $\MstR$, and $M_t^{LR}$.
This yields the tree--level mass relation 
\begin{equation}
m^2_{\tilde{b}_1} = \cos^2\theta_{\tilde t} m^2_{\tilde{t}_1}
 + \sin^2\theta_{\tilde t} m^2_{\tilde{t}_2} - m_t^2 -
 \cos 2\beta \, M_W^2.
\label{eq:msbtree}
\end{equation}

When higher-order corrections are included, the quantities 
$\MstL$ and $\MsbL$ are renormalized by
different counterterms once the on--shell renormalization for the
$\Stop$ and $\Sbot$ squarks is performed. Requiring the SU(2) relation
\BE
\MstL^2 + \de\MstL^2 = \MsbL^2 + \de\MsbL^2
\EE
for the bare parameters at the one--loop level, one obtains $\MstL
\neq \MsbL$ for the renormalized parameters (see also the discussion in
Ref.~\cite{vienna}). 
The difference $\de\MstL^2 - \de\MsbL^2 = \De\msbe^2$ with
\BEA 
\label{eq:msbshA}
\de\MstL^2 &=& \cos^2\tst \de\mste^2 + \sin^2\tst \de\mstz^2
                    + (\mstz^2 - \mste^2) \sin 2 \tst \de\tst 
                    - 2 \mt \de\mt \\
\de\MsbL^2 &=& \de\msbe^2 
\label{eq:msbsh}
\EEA
constitutes a finite $\oas$ contribution to the $\Sbot_1$ mass compared to
its tree--level value. The mass shift $\De\msbe^2$ can also be obtained
by replacing the tree--level quantities in eq.~(\ref{eq:msbtree}) by
their renormalized values and the corresponding counterterms.
We have explicitly checked that the quantity $\De\msbe^2$ is indeed
finite. Note that the squark mass and mixing angle counterterms in
eq.~(\ref{eq:msbshA}--\ref{eq:msbsh}) 
also receive contributions from the pure scalar
diagrams of the type of Fig.~\ref{massct}c which in the two--loop
results given above canceled between the counterterm graphs and the
two--loop diagrams. For the top mass counterterm in
eq.~(\ref{eq:msbsh}) also the graph with gluon exchange enters, which
was absent in the results given above. Due to the latter contribution
the ${\cal O}(\alpha_s)$ correction to eq.~(\ref{eq:msbtree}) is 
different in the dimensional regularization and the dimensional
reduction scheme, while for all results given above the two schemes
yield identical results. In our numerical analysis in this section we
use the result for the top mass counterterm in dimensional reduction.

For the two--loop contribution to the $\rho$--parameter, the
difference generated be using the tree--level mass relations instead
of the one--loop corrected ones is of three--loop order and therefore
not relevant at the order of perturbation under
consideration. However, if the one--loop contribution to $\De\rho$ is
evaluated by calculating the squark masses from the soft breaking
parameters as described above, the $\oas$ correction to the $\Sbot_1$
mass should be taken into account. This gives rise to an extra
contribution compared to the results discussed in section~2.

The tree--level relation $\MstL = \MstR$, which has been widely used in
our numerical examples, can in principle be maintained also for the
renormalized parameters at the one--loop level with on--shell
$\Stop_1, \Stop_2$ mass renormalization. Alternatively, in the spirit
of the discussion given above, one may assume that the symmetry is
extended to the bare parameters:
\BE
\MstL^2 + \de\MstL^2 = \MstR^2 + \de\MstR^2.
\EE
Accordingly, the squark masses and the mixing angle are in this
scenario given in terms of the two parameters $m_{\sq}$ and $M_t^{LR}$,
and there exist two relations between the squark masses and the stop
mixing angle at tree--level,
\begin{eqnarray}
m^2_{\tilde{t}_2} &=& m^2_{\tilde{t}_1} +
  \frac{1}{\sin^2\theta_{\tilde t} - \cos^2\theta_{\tilde t} }
  \cos 2\beta \, M_Z^2 \left(\frac{1}{2} - \frac{4}{3} s_W^2\right),
  \label{eq:mrel1}\\
m^2_{\tilde{b}_1} &=& m^2_{\tilde{t}_1} - m_t^2 +
  \frac{1}{\sin^2\theta_{\tilde t} - \cos^2\theta_{\tilde t} }
  \cos 2\beta \, M_Z^2 \left(c_W^2 - \sin^2\theta_{\tilde t}
  \left(\frac{3}{2} - \frac{2}{3} s_W^2\right) \right) .
  \label{eq:mrel2}
\end{eqnarray}
At ${\cal O}(\alpha_s)$, the right-hand sides of eqs.~(\ref{eq:mrel1})
and (\ref{eq:mrel2}) receive the extra
contributions $\Delta m^2_{\tilde{t}_2}$ and
$\Delta m^2_{\tilde{b}_1}$, respectively, where
\BEA
\De\mstz^2 &=& \de\mste^2 - \de\mstz^2
             + \frac{\sin 2 \tst \de\tst}{(\sin^2\tst - \cos^2\tst)^2}
              \CZb M_Z^2  (-1 + \frac{8}{3} \sw^2), \label{eq:mrel1a}\\
\De\msbe^2 &=& \de\mste^2 - \de\msbe^2 - 2 \mt \de\mt
             + \frac{\sin 2 \tst \de\tst}{2 (\sin^2\tst -
\cos^2\tst)^2}
               \CZb M_Z^2  (-1 + \frac{8}{3} \sw^2).
  \label{eq:mrel2a}
\EEA
These relations have been used in the numerical evaluation displayed
in Fig.~\ref{su2gut.ps}. The difference to the case
with $\MstL = \MstR$ at the renormalized level is only marginal.

The effect of the ${\cal O}(\alpha_s)$ corrections to the mass relations
on $\De\rho_0$ is shown in Fig.~\ref{su2gut.ps}, where the one--loop
correction to the $\rho$ parameter is expressed in terms of $m_{\sq}$,
but with the corrections eqs.~(\ref{eq:mrel1a}--\ref{eq:mrel2a}) to the
mass relations taken into account. The results are shown for the
no--mixing and the maximal--mixing case and for two values of the gluino 
mass, $m_{\tilde{g}} =200$ and 500 GeV. Compared to
Fig~\ref{oneloopdrhoa}, where the tree--level mass relations have been
used as input, the contribution of the $(\tilde{t},\tilde{b})$ doublet
to the $\rho$ parameter is reduced for the parameter space chosen in the
figure. 

The inclusion of the one--loop corrections to the squark mass relations
does not always lead to a decrease of $\De\rho_0$, but can also give
rise to a significant enhancement. This is quantitatively shown in 
Table~2 for the same scenario as in Table 1, 
i.e.\ for the maximal-mixing case with $\Tb = 1$. 
The shift for the sbottom mass in this case follows from 
eqs.~(\ref{eq:msbshA}--\ref{eq:msbsh}) in the limit $\tst = - \pi/4$
with $\de\tst = 0$.
The numerical results are very similar to the case with $\Tb = 1.6$.
The value chosen for the gluino mass is $\mgl=500$ GeV, and for
completeness also the full two--loop contribution $\De\rho_1^{\rm
SUSY}$ is given.
One can see that for large values of the non--diagonal element in the
$\Stop$ mass matrix, $\De\rho_0$ can become larger compared to the 
entries in Table 1, thus significantly 
increasing $\Delta \rho$ at the two--loop level
in a range where the one--loop contribution  is already
quite large. \s

\begin{table}[hbt]
\renewcommand{\arraystretch}{1.5}
\begin{center}
\begin{tabular}{|c||c|c|c|c|} \hline
$m_{\tilde{q}}$ (GeV) & $M_t^{LR}$ (GeV) & $\Delta \rho_0^{\rm SUSY} 
\times 10^{-3}$ & $\De\rho_{1, {\mathrm gluon}}^{\rm SUSY} \times 10^{-4}$ &
 $\De\rho_{1, {\mathrm gluino}}^{\rm SUSY} \times 10^{-4}$ \\ \hline \hline
200 & 100 & 1.17 & 1.92 & 0.09 \\ 
    & 200 & 0.90 & 1.53 & 0.61 \\
    & 300 & 0.57 & 1.05 & 1.37 \\
    & 400 & 1.22 & 1.71 & 2.80 \\ \hline
500 & 200 & 0.23 & 0.39 & -0.08 \\
    & 500 & 0.13 & 0.27 & 0.52 \\
    & 1400& 2.50 & 2.68 & 3.94 \\
    & 1550& 5.71 & 6.18 & 5.34 \\ \hline 
800 & 500 & 0.07 & 0.14 & 0.13 \\
    & 1500& 0.07 & 0.11 & 0.99 \\
    & 2000& 0.31 & 0.42 & 2.31 \\
    & 3700& 14.23 & 16.89 & 16.05 \\ \hline  
\end{tabular}
\renewcommand{\arraystretch}{1.2}
\caption[]{\small $\Delta \rho_0^{\rm SUSY}$ in units of $10^{-3}$
for the same scenario as in Table 1, i.e.\ with $\Tb = 1$, 
for several values of $m_{\tilde{q}}$ and $M_t^{LR}$. The $\oas$
correction to the tree--level mass relation has been
taken into account for $\mgl=500$ GeV. The two--loop contributions
$\De\rho_{1, {\mathrm gluon}}^{\rm SUSY}$ and $\De\rho_{1, {\mathrm
gluino}}^{\rm SUSY}$ are also
given (in units of~$10^{-4}$, $\mgl=500$ GeV).}
\end{center}
\end{table}

In the situation where the relation $\MstL = \MstR$ is relaxed and the
squark masses and the mixing angle are derived from $\MstL, \MstR$ and
$M_t^{LR}$, Fig.~\ref{su2nogutlog.ps} shows
$\De\rho_0^{\rm SUSY}$ for the two choices $\MstL/\MstR = 300/1000$
and $1000/300$ as a function of $M_t^{LR}$. 
For the $\MstL/\MstR = 300/1000$ case 
the solid line corresponds to the use of the tree--level masses, 
while the others reflect the use of the one--loop correction
$\De\msbe^2$ to the tree--level masses following from 
eq.~(\ref{eq:msbshA}--\ref{eq:msbsh}) 
for two gluino masses, $\mgl = 200, 500$ GeV. For the 
$\MstL/\MstR = 1000/300$ case the result is given by the dotted
line. It is insensitive to the one--loop correction to the squark
masses; we therefore show only a single curve. 
For the $\MstL/\MstR = 300/1000$ case, $\De\rho_0$ is decreased
for small values of $M_t^{LR}$ but is increased for large $M_t^{LR}$. 
The effect is more
pronounced for heavier gluinos. For large values of $M_t^{LR}$,
$\De\rho_0$ can become huge in this scenario, exceeding the level of
experimental observability. The bounds on the breaking parameters obtained
from experiment will therefore crucially depend on the proper inclusion
of the two--loop contributions.

\subsection*{5. Conclusions}

We have calculated the ${\cal O}(\alpha_s)$ correction to the squark
loop contributions to the $\rho$~parameter in the MSSM. The result
can be divided into a gluonic contribution, which is typically
of ${\cal O}(10\%)$
and dominates in most of the parameter space, and a gluino contribution,
which goes to zero for large gluino masses as a consequence of decoupling.
Only for gluino and stop/sbottom masses close to their lower experimental
bounds, the gluino contribution becomes comparable to the gluon correction.
In this case, the gluon and gluino contributions add up to $\sim 30\%$ of
the one--loop value for maximal mixing. In general the sum of gluonic and
gluino corrections enters with the same sign as the one--loop contribution.
It thus leads to an enhancement of the one--loop contribution 
(expressed in terms of the physical squark masses) and an
increased sensitivity in the search for scalar quarks through their
virtual effects in high--precision electroweak observables. This is in
contrast to what happens in the SM, where the two--loop QCD corrections
enter with opposite sign and screen the one--loop result. \s

While the gluonic contribution can be presented in a very compact form,
the complete analytical result for the gluino correction is very 
lengthy. We have therefore not written it out explicitly but have given
expressions for three limiting cases, namely the result for zero 
squark mixing, for vanishing gluino mass, and an expansion for a heavy
gluino mass. These limiting cases approximate the exact result
sufficiently well for practical purposes. 
The results have been given in terms of the on--shell squark masses and
are therefore independent of any specific scenario assumed for the mass
values. For different scenarios we have analyzed the extra contributions
caused by the ${\cal O}(\alpha_s)$ correction to the tree--level mass
relations.

\bigskip

\subsection*{Acknowledgements}
We thank A.~Bartl, S.~Bauberger, H.~Eberl and W.~Majerotto for useful 
discussions. A.D. and W.H. would like to thank the Max--Planck 
Institut f\"ur Physik in Munich for the hospitality offered 
to them during the final stage of this work.

\newpage

\input figs.tex

\end{document}

%% file: delrhomglnull.tex
\beqn
\lefteqn{
\Delta\rho^\SU_{1, {\mathrm gluino}}\Bigr|_{\Mgl = 0} = 
      - \frac{\alpha_s}{\pi} \,
               \frac{C_F N_C G_F}{16 \sqrt {2} \pi^2} \Bigg[ \non } \\
  && {} {\rm Li}_2 \KL 1 - \frac{\mt^2}{\msbl^2} \KR 
      \frac{d_{Lt}^2}
           {d_{L1}^2 d_{L2}^2 \mt^2} 
      \bigg( d_{L2}^2 (\msbl^4 - \mste^4 + 2 \msbl^2 \mt^2) \non \\
  && {}         -2 \stt^2 \msbl^2 d_{12} 
              ((\msbl^2 + \mt^2) (\msbl^2 + \mt^2 - \mste^2 - \mstz^2) -
               \mt^4 + \mste^2 \mstz^2) \bigg) \non \\
  && {} + {\rm Li}_2 \KL 1 - \frac{\mt^2}{\mste^2} \KR
      \frac{1}{d_{L1}^2 d_{12} \mt^2} 
      \bigg( -d_{12} (2 \mste^6 (\msbl^2 + \mt^2 - \mste^2)\non \\
  && {}          +2 \msbl^2 \mt^2 (\msbl^2 \mste^2 - \mste^4 + \mt^4)
               - \mt^4 (\mste^4 + 3 \msbl^4))
            + \stt^2 \Big( \mste^4 (\msbl^4 - \mste^4)d_{12}\non \\
  && {}            +2 \mt^2 \mste^2 (\mste^2 \msbl^2 
                                 (\msbl^2 - \mste^2 + 3 \mstz^2)
                                - 2 \mstz^2(\mste^4 + \msbl^4) + \mste^6)\non
\\
  && {}           +\mt^4 (3 \msbl^4 + \mste^4) d_{21} +
                 2 \mt^6 (d_{L1}^2 
                          +\msbl^2 d_{12})\Big)\non \\
  && {}       + 2 \stt^4 \mt^2 \Big( d_{L1}^2 
                                 (\mste^2 \mstz^2 - \mt^4) \Big) \bigg)\non
\\
  && {} + {\rm Li}_2 \KL 1 - \frac{\mt^2}{\mstz^2} \KR
      \frac{1}{d_{L2}^2 d_{12} \mt^2} 
      \bigg( \mstz^4 d_{L2}^2 d_{12} \non \\
  && {}        + \stt^2 \Big( \mstz^4 (\msbl^4 - \mstz^4)d_{21}
                +2 \mt^2 \mstz^2 (\mstz^2 \msbl^2 
                                  (\msbl^2 - \mste^2 - \mstz^2)
                                + \mstz^6)\non \\
  && {}           +\mt^4 (3 \msbl^4 + \mstz^4) d_{12} +
                 2 \mt^6 (-d_{L2}^2 
                          +\msbl^2 d_{21})\Big)\non \\
  && {}        - 2 \stt^4 \mt^2 \Big(d_{L2}^2 
                                 (\mste^2 \mstz^2 - \mt^4) \Big) \bigg)\non
\\
  && {} + \frac{\ln^2(\msbl^2)}
           {2 d_{L1}^2 d_{L2}^2}
      \bigg(  d_{L2}^2 
              (2 \msbl^4 \mste^2 + 2 \msbl^2 \mste^4 \non \\
  && {}          -\mt^2 (3 \msbl^4 + \mste^4) + 2 \msbl^2 \mt^4) 
          - 2 \stt^2 \msbl^2 d_{12} 
            (( \msbl^4 (\msbl^2 + \mste^2 + \mstz^2) \non \\
  && {}          -3 \msbl^2 \mste^2 \mstz^2)
             +\mt^2 (\msbl^2 (\mste^2 + \mstz^2 - 3 \msbl^2) + \mste^2 \mstz^2)
             +\mt^4 (2 \msbl^2 - \mste^2 - \mstz^2)) \bigg) \non \\
  && {} + \frac{-2 \ln(\msbl^2) \ln(\mste^2)}{d_{L1}^2}
      \msbl^2 \mste^4 \ctt^2
     + \frac{-2 \ln(\msbl^2) \ln(\mstz^2)}{d_{L2}^2}
       \msbl^2 \mstz^4 \stt^2 \non \\
  && {} + \frac{- \mt^2 \ln(\msbl^2) \ln(\mt^2)}
           {\mste^2 \mstz^2 d_{L1} d_{L2}}
       \Big( \mstz^2 d_{L2} 
                    (\msbl^2 \mste^2 + \mste^4 - 2 \msbl^2 \mt^2)\non \\
  && {}                -2 \stt^2 \msbl^2 d_{12}
                     (\mt^2 (\msbl^2 - \mste^2 - \mstz^2) 
                             + \mste^2 \mstz^2)\Big)\non \\
  && {} - \frac{2 \ln(\msbl^2) \ln(|d_{1t}|)}
           {\mste^2 d_{L1}^2}
           \msbl^4 d_{1t}^2 \ctt^2
   -  \frac{2 \ln(\msbl^2) \ln(|d_{2t}|)}
           {\mstz^2 d_{L2}^2}
           \msbl^4 d_{2t}^2 \stt^2 \non \\
  && {} + \frac{ \msbl^2 \ln(\msbl^2)}
           {d_{L1} d_{L2}}
       \Big( (\msbl^2 - 3 \mste^2) d_{L2}
              +2 \msbl^2 d_{12} \stt^2 \Big) \non \\
  && {} - \frac{\ctt^2 \ln^2(\mste^2)}
           {2 d_{L1}^2 d_{12}}
       \bigg(d_{12} 
            (2 \msbl^4 \mste^2 - 2 \msbl^2 \mste^4 - 3 \msbl^4 \mt^2 
             - \mste^4 \mt^2 \non \\
  && {}        + 2 \msbl^2 \mt^4) 
            + 2 d_{L1}^2 
                (\mste^2 \mstz^2 - \mt^4) \stt^2 \bigg)\non \\
  && {} + \frac{ \ctt^2 \mt^2 \ln(\mste^2) \ln(\mt^2)}
           {\mste^2 \mstz^2 d_{L1} d_{12}}
      \Big(\mstz^2 d_{12}
                   (\msbl^2 \mste^2 + \mste^4 - 2 \msbl^2 \mt^2) \non \\
  && {}       +2 \stt^2 d_{L1} (\mste^2 + \mstz^2) 
               (2 \mste^2 \mstz^2 - \mste^2 \mt^2 - \mstz^2 \mt^2) \Big)\non
\\
  && {} + \frac{2 d_{1t}^2 \ln(\mste^2) 
            \ln(|d_{1t}|) \ctt^2}
           {\mste^2 d_{L1}^2 d_{12}^2}
       \Big( \msbl^4  d_{12}^2
             - d_{L1}^2 \mstz^4 \stt^2 \Big)\non \\
  && {} + \frac{2 d_{2t}^2 \ln(\mste^2) \ln(|d_{2t}|)}
           { \mstz^2 d_{12}^2} \stt^2 \ctt^2 \mste^4 \non \\
  && {} - \frac{ \ln(\mste^2) }
           {2 d_{L1} d_{12} \mt^2}
       \Big(d_{12} (\mste^4 d_{L1} + 4 \mt^2 \mste^2 d_{1L}
               -4 \msbl^2 \mt^4) \non \\
  && {}         - 2 \stt^2 \mt^2 ( \mste^4 (\msbl^2 - \mste^2 - \mstz^2)
                              +\msbl^2 \mste^2 \mstz^2 
                            +\mt^2 ( 2 (\mste^2 + \mstz^2) d_{L1} 
                                     -2 \mste^4 \non \\
  && {}                                + 2 \msbl^2 \mstz^2))
              +2 \stt^4 \mt^2 d_{L1}
                 (\mste^2 (\mste^2 + \mstz^2) 
                  + 2 \mt^2 (2 \mste^2 + \mstz^2)) \Big) \non \\
  && {} + \frac{ \ln^2(\mstz^2) \stt^2}
           {2 d_{L2}^2 d_{12}}
       \Big(  2 \mstz^4 d_{1L} d_{2L}
            +\mt^2 d_{12} (3 \msbl^4 + \mstz^4) \non \\
  && {}       -2 \mt^4 (d_{L2}^2 + \msbl^2 d_{12})
            -2 \stt^2 d_{L2}^2 (\mste^2 \mstz^2 - \mt^4) \Big)\non \\
  && {} - \frac{ \mt^2 \stt^2 \ln(\mstz^2) \ln(\mt^2)}
           {\mste^2 \mstz^2 d_{L2} d_{12}}
      \Big(\mstz^2 (\mste^2 ( 3 \msbl^2 (\mste^2 + \mstz^2) 
                            +\mstz^2 (2 \msbl^2 - 3 \mstz^2 - 5 \mste^2))\non
\\
  && {}               + 2 \mt^2 ((\mste^2 + \mstz^2)^2 
                               - \msbl^2(3 \mste^2 + \mstz^2)))
            -2 \stt^2 d_{L2} (\mste^2 + \mstz^2) \non \\
  && {}                (2 \mste^2 \mstz^2 - \mt^2 (\mste^2 + \mstz^2)) \Big)\non
\\
  && {} + \frac{2 \mstz^4 d_{1t}^2 \stt^2 \ctt^2
            \ln(\mstz^2) \ln(|d_{1t}|)}
           {\mste^2 d_{12}^2} \non \\
  && {} - \frac{2 d_{2t}^2 \stt^2 \ln(\mstz^2) 
             \ln(|d_{2t}|)}
           {\mstz^2 d_{L2}^2 d_{12}^2}
        \Big( \mstz^2 d_{L1} 
                    (\msbl^2d_{12} + \mste^2 d_{L2})
             - \mste^4 d_{L2}^2 \stt^2 \Big)\non \\
  && {} - \frac{ \ln(\mstz^2)}
           {2 d_{L2} d_{12} \mt^2}
       \Big( \mstz^2 d_{L2} d_{12} (\mstz^2 -4 \mt^2)
               + 2 \stt^2 \mstz^2 \mt^2 \non \\
  && {}             (d_{L2} (\mste^2 + \mstz^2) 
                   +\mt^2 (6 \msbl^2 - 2 \mste^2 - 4 \mstz^2)) \non \\
  && {}          - 2 \stt^4 d_{L2} \mt^2 
                  (\mstz^2 (\mste^2 + \mstz^2) 
                   + 2 \mt^2 (\mste^2 + 2 \mstz^2)) \Big) \non \\
  && {} - \frac{ \ln(\mt^2)}
           {\mste^2 \mstz^2 d_{L1} d_{L2}}
      \bigg( \Big( \mstz^2 d_{L2} 
          ( \mste^2 d_{L1} (\msbl^2 + 2 \mste^2 + \mstz^2)\non \\
  && {}      + \mt^2 \mste^2 (\mste^2 - 3 \msbl^2)
           + \mt^4 (3 \msbl^2 - \mste^2)) \Big)
            - \stt^2 \mstz^2 \Big( 
                    \mste^2 d_{1L} d_{L2} d_{21} \non \\
  && {}    +2 \mt^2 \mste^2 (-5 \msbl^4 + 4 \msbl^2 \mste^2
                                 +6 \msbl^2 \mstz^2 - 5 \mste^2 \mstz^2)\non
\\
  && {}            + \mt^4 (6 \msbl^2 (\msbl^2 - \mste^2 - \mstz^2)
                        +2 \mste^2 (\mste^2 + 2 \mstz^2)) \Big) \non \\
  && {}       - \stt^4 \Big( d_{L1} d_{L2} \mt^2 
                  (10 \mste^2 \mstz^2 - 3 \mt^2 (\mste^2 + \mstz^2)) 
             \Big) \bigg)\non \\
  && {} + \frac{ d_{1t}^2 \ln(|d_{1t}|)}
           {2 \mste^2 \mt^2 d_{L1} d_{12}}
       \Big( d_{12} (\mste^2 d_{L1} 
                             + 2 \mt^2 (3 \msbl^2 - \mste^2)) \non \\
  && {}     - 4 \stt^2 \mt^2 ( -d_{1L}^2 + d_{L2}^2
                           - \mstz^2 (\msbl^2 - 2 \mste^2 + \mstz^2)) 
          + 2 \stt^4 \mt^2 d_{L1} (\mste^2 - 3 \mstz^2) \Big)\non \\
  && {} + \frac{ d_{2t}^2 \ln(|d_{2t}|)}
           {2 \mstz^2 \mt^2 d_{L2} d_{12}}
        \Big( \mstz^2 d_{L2} d_{12}
          - 4 \stt^2 \mt^2 \mstz^2 d_{L1} 
          + 2 \stt^4 \mt^2 d_{L2} (3 \mste^2 - \mstz^2) \Big)\non \\
  && {} + \frac{1}{6 \mt^2}
       \bigg( \Big( 21 \mt^2 (\mste^2 + \mstz^2) + 12 \mt^2 d_{L2}
             -\pi^2 ((\msbl^2 + \mste^2)^2 + \mste^4 + \mstz^4 
                      -2 \mste^2 \mt^2) \non \\
  && {}                 - 6 \mt^4 \Big)
             + \stt^2 \Big( 6 \mt^2 (-3 \mste^2 + \mstz^2 - 6 \mt^2)
                     + \pi^2 ((\mste^2 + \msbl^2)^2 - (\mstz^2 + \msbl^2)^2\non \\
  && {}                        - 2 \mt^2 d_{12}) \Big)
             + 6 \stt^4 \mt^2 (\mste^2 + \mstz^2 + 6 \mt^2) \bigg) \Bigg] .
\label{eq:lightgluino}
\eeqn

%% file: delrhomglseries.tex
\beqn
\lefteqn{
\Delta\rho^\SU_{1, {\mathrm gluino}} = 
\frac{\alpha_s}{\pi} \,
               \frac{C_F N_C G_F}{16 \sqrt {2} \pi^2} \Bigg\{ \non } \\
&& {}       -2\, \frac{\mt}{\Mgl} 
               \,\frac{ \stt\, \ctt\, (1-2 \stt^2)}{d_{L1} \,d_{L2}\, d_{12}}
 \ \Bigg[           d_{L1} d_{L2} d_{12}
          \left(\mste^2 \ctt^2 - \mstz^2 \stt^2\right)   
  -  \msbl^4 d_{12}^2 \ln \msbl^2\non\\ && {}
         + \mste^2 d_{L2} \ln \mste^2 \left(\msbl^2 \mste^2 
                   -2 \ctt^2 \mstz^2 \msbl^2 
    + (1-2 \stt^2) \mste^2 \mstz^2\right) \non\\
 && {}
         +  \mstz^2 d_{L1} \ln\mstz^2
                       \left(\msbl^2 (\mstz^2 - 2 \stt^2 \mste^2) 
                        - (1-2 \stt^2) \mste^2 \mstz^2\right)   \Bigg] \non\\
  && {} + \frac{1}{\Mgl^2}\,
      \frac{1}{3 d_{L1} \,d_{L2}\,d_{12}} \ \Bigg[ 2\,d_{L1}\, d_{L2} \,d_{12}
              \left[d_{L1}^2 - \stt^2 ( d_{12}^2 \ctt^2 + d_{1L}^2 -d_{2L}^2) 
              \right. \non\\
  && {}
 +\left. 3 \mt^2 \left(-\mste^2 (1+2 \stt^2) + 2 \msbl^2 - 4 \mstz^2 \stt^2
                         + 3 \stt^4 (\mste^2 + \mstz^2) \right) \right] \non\\
  && {}  \     -  6 \,\msbl^4 \,d_{12} \,\mt^2 
             \ln\msbl^2    \left(d_{L2} - \stt^2 d_{12}\right) \non\\
  && {}   \    + 6 \,\mste^2 d_{L2} \mt^2 \ctt^2  \ln\mste^2 
                 \left( d_{12} \,( 2\msbl^2- \mste^2)
                + 6\, \stt^2 \,  d_{L1}\,   \mstz^2 \right)\non\\
  && {}    \   + 6 \, \mstz^2\,  d_{L1} \, \mt^2 \, \stt^2 \, \ln\mstz^2
                 \left(
        \mstz^4-4 \msbl^2 \mste^2 - 2 \msbl^2 \mstz^2 + 5 \mste^2 \mstz^2 
                  + 6 \, \stt^2\,  d_{L2}\, \mste^2  \right) \non\\
  && {}     \  +  3 \,d_{L1} \,d_{L2}\,    d_{12} \,\mt^2 \,\ln \mt^2  
 \left(\msbl^2 + \mste^2 - 2 \mt^2 - \stt^2 (5 \mste^2 + 3 \mstz^2)
                  +4 \stt^4 (\mste^2 + \mstz^2)\right) \non\\
  && {}      \ - 6 \,\msbl^2 \,d_{12}\, \mt^2 \ln\msbl^2 \ln(\mt^2/ \Mgl^2) 
                 \left(\mste^2 d_{L2} - \stt^2 \,\msbl^2 \,d_{12}\right) \non\\
  && {}     \  + 6 \, \mste^2 d_{L2} \mt^2 \ctt^2
        \ln\mste^2 \ln(\mt^2/\Mgl^2)       
           \left(\msbl^2 d_{12} + 4 \stt^2 \mstz^2 d_{L1}\right) \non\\
  && {}     \  +    6 \, d_{L1} \, \mstz^2 \, \mt^2 \, \stt^2 
       \, \ln\mstz^2 \, \ln(\mt^2/\Mgl^2)           \left(\msbl^2 d_{12} 
                  + 4 \ctt^2 \mste^2 d_{2L}\right) \non\\
  && {}     \   +3 \, \mt^2 \, 
               d_{L1} \,  d_{L2} \,  d_{12}  \, \ln\Mgl^2 \left(
               \msbl^2 - 3 \mste^2 + 2 \mt^2 
              + 4 \stt^2 \ctt^2 (\mste^2+\mstz^2) + 3 \stt^2 d_{12})\right) 
        \Bigg] \non\\
&& {} + \frac{\mt}{\Mgl^3}\, \frac{\stt\,\ctt}{6 \, d_{12} \,d_{L1}\,d_{L2}}
    \Bigg[
    12\,{  d_{L1}}\,{  d_{L2}}\,{  d_{12}^2}\,\mt^2 \,{\pi^2}
  +   36\,{  d_{L1}}\,{  d_{L2}}\,{  d_{12}^2}\,\mt^2 \,{{\ln^2   \Mgl^2}}\non\\
&& {}  - {  d_{L1}}\,{  d_{L2}}\,{  d_{12}}\,
   \Big( 4\,\msbl^2\,\mste^2 + 4\,\mste^4 - 
     4\,\msbl^2\,\mstz^2 + 4\,\mste^2\,\mstz^2 + 
     57\,\mste^2\,\mt^2  - 21\,\mstz^2\,\mt^2  \non\\
&& {} - 20\,\mste^4\,{\stt^2}  - 16\,\mste^2\,\mstz^2\,{\stt^2} + 
     4\,\mstz^4\,{\stt^2} - 
     108\,\mste^2\,\mt^2 \,{\stt^2} - 
     36\,\mstz^2\,\mt^2 \,{\stt^2} + 
     16\,{\mste^4}\,{\stt^4} \non\\
&& {} + 16\,\mste^2\,\mstz^2\,{\stt^4} + 72\,\mste^2\,\mt^2 \,{\stt^4} + 
     72\,\mstz^2\,\mt^2 \,{\stt^4} \Big)  \non\\
&& {} -  24\,{  d_{L1}}\, d_{L2}\,  d_{12}\,\mt^2 \ln {  \Mgl^2}\,
   \Big( 3\,\mste^2 - 4\,\mstz^2 + 3\,\mste^2\,{\stt^2} + 
     \mstz^2\,{\stt^2} - 2\,\mste^2\,{\stt^4} - 
     2\,\mstz^2\,{\stt^4} \Big)  \non\\
&& {}
 +   4\,{  d_{12}^2}\,{\msbl^4}\,\ln \msbl^2
   \Big( 2\,\mste^2 + 15\,\mt^2  - 4\,\mste^2\,{\stt^2} - 
     18\,\mt^2 \,{\stt^2} \Big)  \non\\
&& {}
   - 4\,{  d_{L2}}\,\mste^2\,\ln \mste^2\Big( 2\,\msbl^2\,\mste^4 - 
     4\,\msbl^2\,\mste^2\,\mstz^2 + 
     2\,\mste^4\,\mstz^2 + 3\,\msbl^2\,\mste^2\,\mt^2  + 
     12\,\mste^4\,\mt^2 \non\\
&& {}  - 12\,\msbl^2\,\mstz^2\,\mt^2  - 
     3\,\mste^2\,\mstz^2\,\mt^2  - 
     4\,\msbl^2\,\mste^4\,{\stt^2} + 
     12\,\msbl^2\,\mste^2\,\mstz^2\,{\stt^2} - 
     8\,\mste^4\,\mstz^2\,{\stt^2} \non\\ 
&& {} - 18\,\msbl^2\,\mste^2\,\mt^2 \,{\stt^2} + 
     54\,\msbl^2\,\mstz^2\,\mt^2 \,{\stt^2} - 
     36\,\mste^2\,\mstz^2\,\mt^2 \,{\stt^2} - 
     8\,\msbl^2\,\mste^2\,\mstz^2\,{\stt^4} \non\\
&& {} + 8\,\mste^4\,\mstz^2\,{\stt^4} - 
     36\,\msbl^2\,\mstz^2\,\mt^2 \,{\stt^4} + 
     36\,\mste^2\,\mstz^2\,\mt^2 \,{\stt^4} \Big) \non\\
&& {}
 + 4\,{  d_{L1}}\,\mstz^2\,\ln \mstz^2
   \Big( -2\,\msbl^2\,\mste^2\,\mstz^2 + 
     2\,\mste^4\,\mstz^2 - 6\,\msbl^2\,\mste^2\,\mt^2  - 
     3\,\msbl^2\,\mstz^2\,\mt^2 \non\\
&& {}  + 21\,\mste^2\,\mstz^2\,\mt^2 - 12\,\mstz^4\,\mt^2  + 
     4\,\msbl^2\,\mste^4\,{\stt^2} + 
     4\,\msbl^2\,\mste^2\,\mstz^2\,{\stt^2} - 
     8\,\mste^4\,\mstz^2\,{\stt^2} \non\\
&& {} + 18\,\msbl^2\,\mste^2\,\mt^2 \,{\stt^2} 
      + 18\,\msbl^2\,\mstz^2\,\mt^2 \,{\stt^2} - 
     36\,\mste^2\,\mstz^2\,\mt^2 \,{\stt^2} - 
     8\,\msbl^2\,\mste^4\,{\stt^4} + 
     8\,\mste^4\,\mstz^2\,{\stt^4} \non\\
&& {} - 36\,\msbl^2\,\mste^2\,\mt^2 \,{\stt^4} + 
     36\,\mste^2\,\mstz^2\,\mt^2 \,{\stt^4} \Big) \non\\
&& {}
- 24\,{  d_{L1}}\,{  d_{L2}}\,{  d_{12}}\,\mt^2 \,\ln \mt^2
   \Big( -\mste^2 + 2\,\mstz^2 - 3\,\mste^2\,{\stt^2} - 
     \mstz^2\,{\stt^2} + 2\,\mste^2\,{\stt^4} + 
     2\,\mstz^2\,{\stt^4} \Big)   \non\\
&& {} 
 - 36\,{  d_{L1}}\,{  d_{L2}}\,{{{  d_{12}}}^2}\,\mt^2 \,\ln {  \Mgl^2}\,
   \ln \mt^2
+ 12\,  d_{12}^2\,\msbl^4\,\mt^2 \,\ln \msbl^2\,\ln (\mt^2/\mgl^2)
   \Big( 3 - 4\,{\stt^2} \Big)   \non\\
&& {}+ 
  12\,{  d_{L2}}\,\mste^2\,\mt^2 \,\ln \mste^2 \ln (\mt^2/\mgl^2) 
   \Big( -3\,\mste^4 + 2\,\msbl^2\,\mstz^2 + 
     \mste^2\,\mstz^2 + 4\,\msbl^2\,\mste^2\,{\stt^2} \non\\
&& {} - 12\,\msbl^2\,\mstz^2\,{\stt^2}
     + 8\,\mste^2\,\mstz^2\,{\stt^2} + 
     8\,\msbl^2\,\mstz^2\,{\stt^4} - 
     8\,\mste^2\,\mstz^2\,{\stt^4} \Big) \non\\
&& {}
- 12\,{  d_{L1}}\,\mstz^2\,\mt^2 \,\ln  \mstz^2 \ln (\mt^2/\mgl^2) 
   \Big( 2\,\msbl^2\,\mste^2 - 5\,\mste^2\,\mstz^2 + 
     3\,\mstz^4 - 4\,\msbl^2\,\mste^2\,{\stt^2} \non\\
&& {} - 4\,\msbl^2\,\mstz^2\,{\stt^2}
    + 8\,\mste^2\,\mstz^2\,{\stt^2} 
    + 8\,\msbl^2\,\mste^2\,{\stt^4} - 
     8\,\mste^2\,\mstz^2\,{\stt^4} \Big) \,
\Bigg] \Bigg\}
+ {\cal O}\left(\frac1{\Mgl^4}\right) .
\label{heavygluino}
\eeqn

%% file: delrhonomix.tex
\beqn
\lefteqn{
\Delta\rho^{\rm SUSY}_{1, {\mathrm gluino}}\Bigr|_{M_t^{LR}=0} =
      \frac{\alpha_s}{\pi} \,
               \frac{C_F N_C G_F}{16 \sqrt {2} \pi^2} \Bigg\{ \non } \\
&& {}  13\mgl^2 - 4\msbl^2 - \frac9{2} \mste^2 - \frac1{2}\mstz^2 -
\frac{(\mste^2-5\msbl^2  )\,\mt^2}{d_{L1}} \non \\ 
&& {} + \frac1{2} \sum_{i=1,2}\left(\mgl^2 - \msti^2 + \mt^2\right) 
     B_0^{\mathrm fin}(\mt^2, \mgl, \msti)  \non \\
&& {} -\frac{\mgl^2-\mste^2+\mt^2       }{d_{L1}^2} 
      B_0^{\mathrm fin}(\mste^2, \mgl, \mt    ) 
      \KL  3\msbl^4 -4 \msbl^2 \mste^2 + \mste^4
      -2 \msbl^4 \ln \frac{\msbl^2}{\mste^2}         \KR \non \\
&& {} -\frac{d_{gL}       }{d_{L1}^2} 
      B_0^{\mathrm fin}(\msbl^2, \mgl, 0    ) 
      \KL \msbl^4-4 \msbl^2 \mste^2 + 3 \mste^4
      -2 \mste^4 \ln\frac{\mste^2}{\msbl^2}        \KR \non \\
  && {} - \frac{d_{gL} } {d_{1L}^2 \,\mt^2} 
      {\rm Li}_2 \KL 1 - \frac{\mgl^2}{\msbl^2} \KR   
      \left(d_{gL} d_{1L} ( \msbl^2-2 \mgl^2  + \mste^2)
           +2 \mste^2 \mt^2 (2 \mste^2 - \msbl^2 - \mgl^2)\right) \non \\
  && {} - \frac{2 d_{g1}^2} {d_{1L}^2 \,\mt^2} 
      {\rm Li}_2 \KL 1 - \frac{\mgl^2}{\mste^2} \KR 
      \left(\mgl^2 d_{1L} + \mste^2 (\msbl^2 - \mste^2 + \mt^2)\right)
  + \frac{d_{2g}^2}{\mt^2} 
       \,{\rm Li}_2 \KL 1 - \frac{\mgl^2}{\mstz^2} \KR \non \\
&& {} + \left(
3\,\mgl^2 + \sum_{i=1,2}{{\mgl^6 - 2\,\mgl^4\,\msti^2 + 
       \mgl^2\,\msti^4 - \mgl^4\,\mt^2  - 
       \mgl^2\,\msti^2\,\mt^2 }\over {\lambda^2(\mgl^2,\mt^2 ,\msti^2)}} 
\right) \ln \mgl^2 \non \\
&& {} - \frac{2}{d_{L1}^2} \KL 
d_{L1} (\mgl^2(3\msbl^2 +\mste^2)- 2 \msbl^2 \mste^2) + 2 \msbl^4 \mt^2
\KR \ln \msbl^2 \non \\
&& {} + \KL
  {{4\,\mgl^2(\msbl^2 + 3\mste^2) - 
       11\msbl^2\mste^2 + 3\mste^4}\over 
     {2\,d_{L1}}} \right.\non \\
&& {} + \left. {{2{\msbl^2}
       ( {\msbl^2} + {\mste^2} ) {\mt^2}}
      \over {{d_{L1}^2}}} - 
   {{2\,{\mgl^2}\,{\mste^2}\,{\mt^2}}\over 
     {\lambda^2({\mgl^2},{\mt^2},{\mste^2})}}
\KR \ln \mste^2\non \\
&& {} - \KL   {{3\,{\mstz^2}}\over 2} + 
   {{2\mgl^2\mstz^2\mt^2}\over      {\lambda^2({\mgl^2},{\mt^2},{\mstz^2})}}
\KR \ln \mstz^2  
+ \frac{d_{g2}^2}{2\mt^2}\ln \mstz^2 \left(\ln\frac{\mstz^2}{\mgl^2}-
\ln\mt^2\right)\non \\
&& {} +
\KL -2\mgl^2 + \msbl^2 + 2\mste^2 + \mstz^2 +
\sum_{i=1,2}\frac{\mgl^2\left(\mt^2
 (3\msti^2+\mgl^2) -d^2_{gi}\right)}{\lambda^2(\mgl^2, \mt^2, \msti^2)} 
\KR \ln\mt^2\non \\
&& {} +\frac{d_{g1}^2}{d_{L1}^2\,\mt^2} \KL
\mgl^2 \, d_{L1} -\mste^2 (d_{L1}+\mt^2) 
\KR \ln^2 \mste^2\non \\
&& {} - \frac{d_{gL}}{2 d_{L1}^2\mt^2} \KL
d_{L1} \, d_{Lg} \,(\msbl^2 +\mste^2-2\mgl^2) +
2 \,\mste^2\mt^2 (2\mste^2-\msbl^2-\mgl^2)
\KR \ln^2\msbl^2\non \\
&& {} -\left[  \frac{\mt^2}{2} + \frac{1}{d_{L1}^2 \mt^2}\left(
\mt^2(\mgl^4 (3\msbl^2-4\mste^2)-\mste^2\,d_{L1}^2 + \mste^4(2\mgl^2-\msbl^2))
+d_{g1}^3 d_{L1} \right.\right.\non \\
&& {} +\left.\left.
\mt^4 ( \msbl^2 (\mste^2+\msbl^2) -\mgl^2 d_{L1}) -\mt^6 \msbl^2
\right)\right] \ln\mste^2\,\ln\mgl^2\non \\
&& {} -\frac1{2 \,d_{L1}^2 \mt^2}
\left( d_{gL}^2 (d_{g1}+d_{gL})
            d_{1L} 
         + 2 \mt^2 (\mgl^4 (-3 \msbl^2 + 4 \mste^2) 
                       + \mste^4 (\msbl^2 - 2 \mgl^2)) \right.\non \\
  && {} +  \left.  \mt^4 ((\msbl^2 + \mgl^2)^2 - (\mste^2 + \mgl^2)^2 - 4 \msbl^4)
            +2 \mt^6 \msbl^2 \right)
 \ln \mgl^2 \ln \msbl^2\non \\
&& {} + \left( 5\,d_{g1}^2 + d_{g2}^2 - 4\,d_{g1}\,d_{L1} +
 d_{L1}^2 - 2\mste^2\mt^2 \right) (\ln \mgl^2 \ln \mt^2)/(2\,\mt^2)\non \\
&& {} + 
\left( d_{gL}^2 (d_{g1}+d_{gL})
              d_{L1} \right.
-   2 \mt^2 (\mgl^2 \msbl^2 (3 \mgl^2 - 2 \msbl^2) 
                       - 2 \mgl^4 \mste^2 + \msbl^2 \mste^4) \non \\
  && {} +      \left.\mt^4 d_{L1} (2 \mgl^2 + \mste^2 + \msbl^2)
            + 2 \mt^6 \msbl^2 \right) \frac{\ln \msbl^2 \ln \mt^2}{2\,\mt^2
d_{L1}^2}\non \\
&& {} + \left[ \frac{\mt^2}{2}-\mste^2 -\frac1{d_{L1}^2 \,\mt^2}
\left(   d_{g1}^3 d_{L1} 
             +\mt^2 d_{g1} (2 d_{1L}^2 
                     +\msbl^2 \mste^2 + \mgl^2 (2 \mste^2 - 3
\right.\right.                     \msbl^2)) \non \\
  && {} +  \left.\left.\mt^4 (\mgl^2 + \msbl^2) d_{L1}
             + \mt^6 \msbl^2 \right)\right]\ln \mt^2\ln\mste^2\non \\
&& {} - \frac{\lambda^2(\mgl^2, \mt^2, \msbl^2)}{2 \mt^2 \msbl^2 \,d_{L1}^2}\left[
 \mt^2 (\mste^2 (\mste^2 + 2 \mgl^2) 
               + \msbl^2 (\msbl^2 - 2 \mt^2 - 4 \mgl^2)) \right.\non \\
  && {} +       \left. d_{gL} d_{L1} (d_{gL}+d_{g1})
             \right] \Phi(\mgl^2,\mt^2,\msbl^2)\non \\
&& {} +
\left(\frac{d_{g2}^3 + d_{g2}^2\mt^2 - 2\mgl^4\mt^2}{2\,\mt^2\,\mstz^2}+
\frac{\mgl^2(d_{g2}^2 + d_{g2}\mt^2 - 4\mgl^2\mt^2)
}{\lambda^2(\mgl^2, \mt^2, \mstz^2)}
\right)\Phi(\mgl^2,\mt^2,\mstz^2)\non \\
&& {} + \left[\frac{d_{g1}^2\mgl^2 + d_{g1} \mgl^2\mt^2 - 4\mgl^4\mt^2}
{\lambda^2(\mgl^2, \mt^2, \mste^2)}\right.\non \\
&& {} +\frac1{2\,\mt^2\mste^2 d_{L1}^2}\left(
  2 \,d_{g1}^4(d_{L1} + \mt^2) + 2d_{g1}^3\mt^2( 3\mt^2-\mgl^2 ) \right.+ 
   2d_{L1}\mt^4 (4d_{g1}\mgl^2 \non \\
&& {} + 4\mgl^4 -
 4d_{g1} \mt^2 + 4\mgl^2\mt^2 - \mt^4) + 
   2d_{g1}^2\mt^2(d_{L1}^2 - 4d_{L1}\mgl^2 - 4d_{L1}\mt^2 
\non \\
&& {} - 7\mgl^2\mt^2 
+ 3\mt^4) + 
   d_{L1}^2\mt^2(4d_{g1}\mgl^2 - 2\mgl^4 + 5d_{g1}\mt^2 -
 4\mgl^2\mt^2 + 3\mt^4)\non \\
&& {} + 
   \left.\left.
2\mt^4(4d_{g1}\mgl^4 - 7d_{g1}\mgl^2\mt^2 + 4\mgl^4\mt^2 + d_{g1}\mt^4 - 
      \mgl^2\mt^4)
\right)
\right] \Phi(\mgl^2,\mt^2,\mste^2)
\Bigg\} .
\eeqn

%% file: figs.tex

\vspace*{6cm}

\begin{figure}[ht]
\begin{center}
\mbox{
\psfig{figure=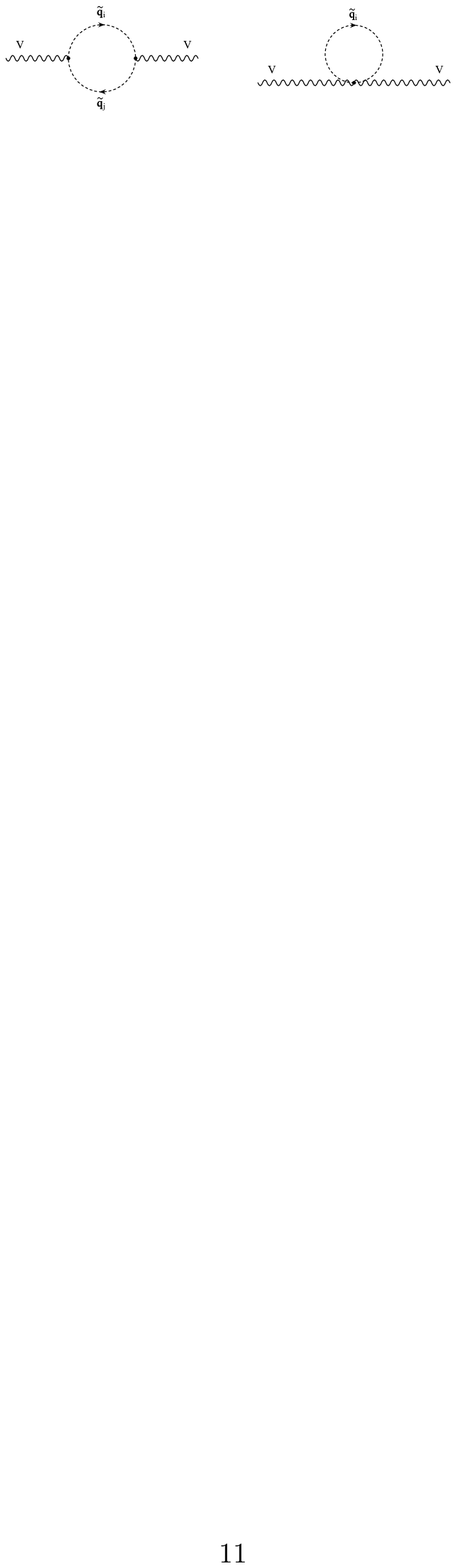,width=12cm,bbllx=210pt,bblly=680pt,bburx=397pt,
bbury=720pt}}
\end{center}
\caption[]{Feynman diagrams for the contribution of scalar quark loops 
to the gauge boson self--energies at one--loop order.}
\label{oneloopdiagrams}
\end{figure}

\begin{figure}[ht]
\begin{center}
\mbox{
\psfig{figure=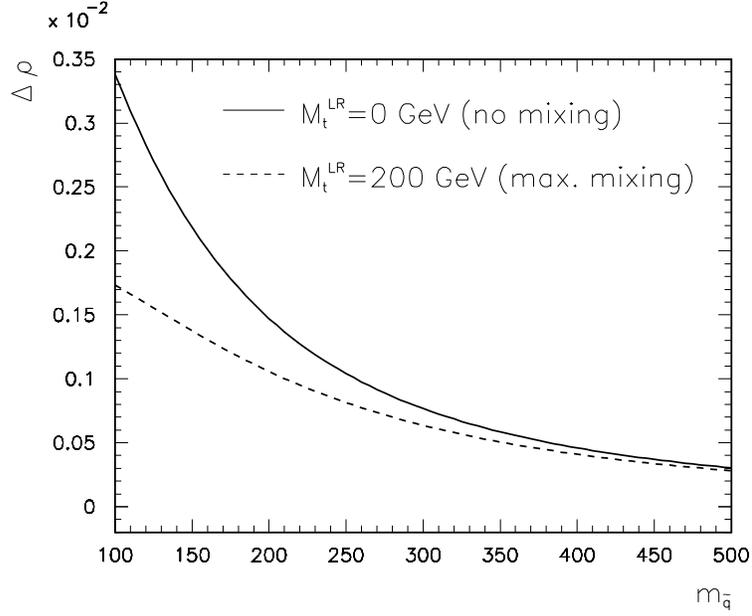,width=10cm,bbllx=120pt,bblly=270pt,%
                                        bburx=450pt,bbury=540pt}
     }
\parbox{15cm}{
\caption{\label{oneloopdrhoa}
One--loop contribution of the $(\tilde t, \tilde b)$ doublet to
$\Delta\rho$ as a function of the common squark mass $\msq$ for
$\theta_{\tilde t} = 0$ and $\theta_{\tilde t} \sim \pi/4$ (with $\Tb
= 1.6$,  $M^{LR}_b = 0$, and  $M^{LR}_t = 0 \mbox{ or } 200  
\mbox{ GeV}$).
        }    }
\end{center}
\end{figure}

\begin{figure}[ht]
\begin{center}
\mbox{
\psfig{figure=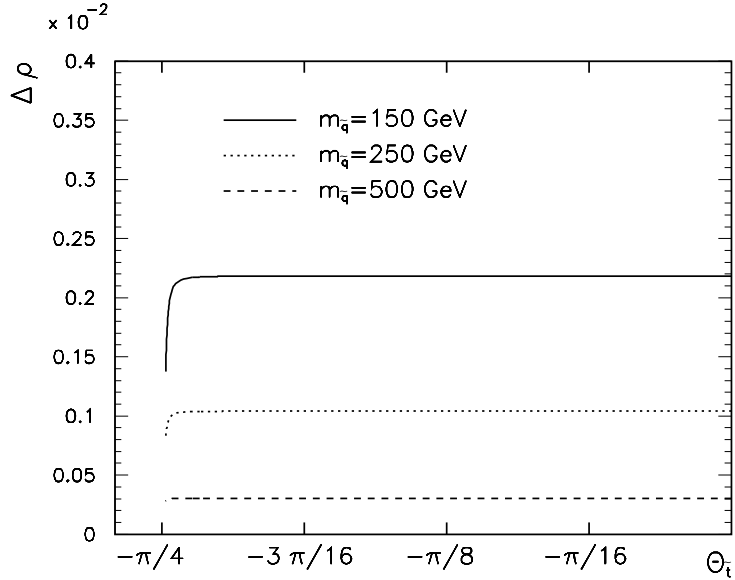,width=10cm,bbllx=120pt,bblly=270pt,%
                                        bburx=450pt,bbury=540pt}
     }
\parbox{15cm}{
\caption{\label{oneloopdrhob}
Dependence of the one--loop contribution $\dr_0^\SU$ on the stop mixing
angle $\theta_{\tilde t}$. The parameters are the same as in 
Fig.~\ref{oneloopdrhoa}.
        }    }
\end{center}
\end{figure}

\begin{figure}[ht]
\begin{center}
\mbox{
\psfig{figure=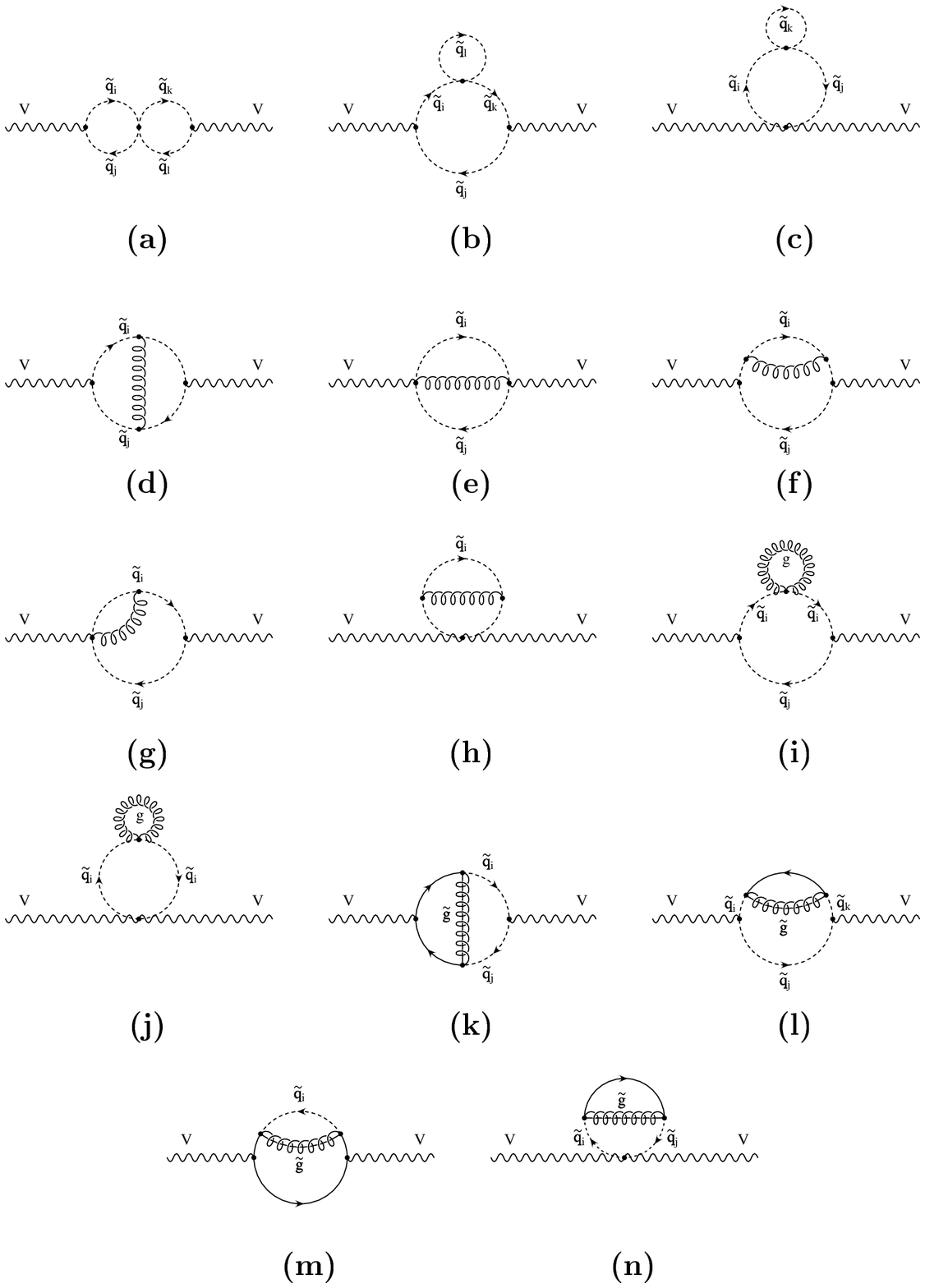,width=12cm,bbllx=90pt,bblly=150pt,%
                                        bburx=510pt,bbury=740pt}}
\end{center}
\caption{Feynman diagrams for the contribution of scalar quark loops 
to the gauge--boson self--energies at two--loop order.}
\label{twoloopdiagrams}
\end{figure}

\begin{figure}[ht]
\begin{center}
\mbox{
\psfig{figure=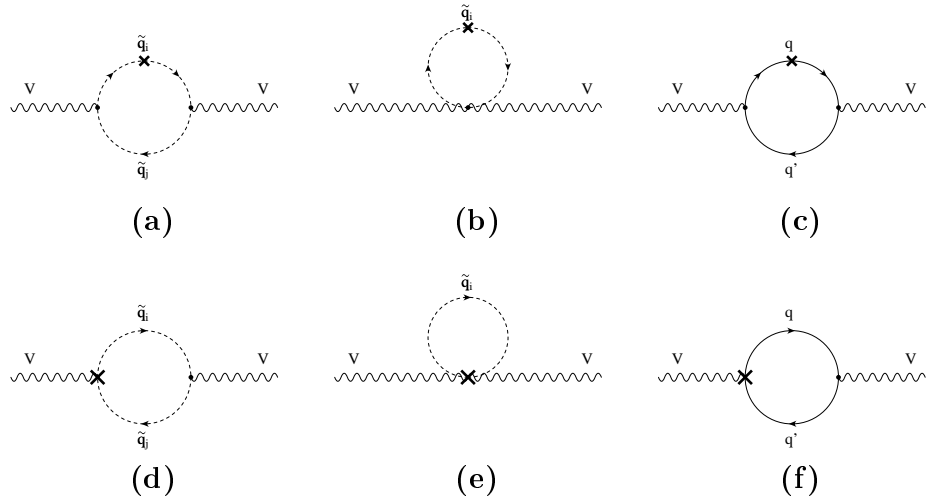,width=12cm,bbllx=90pt,bblly=510pt,%
                                   bburx=510pt,bbury=740pt}}
\end{center}
\caption{Counterterm contributions to the gauge--boson self--energies at 
two--loop order.}
\label{twoloopcts}
\end{figure}

\begin{figure}[ht]
\begin{center}
\mbox{
\psfig{figure=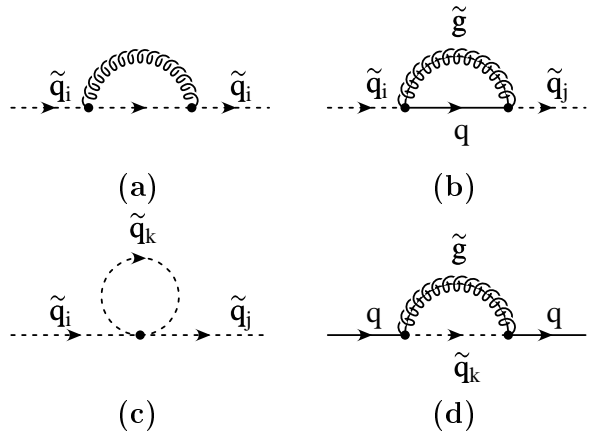,width=12cm,bbllx=90pt,bblly=530pt,%
                                   bburx=520pt,bbury=750pt}}
\end{center}
\caption{One--loop diagrams contributing to the squark mass and 
         mixing angle counterterms.}
\label{massct}
\end{figure}


\begin{figure}[ht]
\begin{center}
\mbox{
\psfig{figure=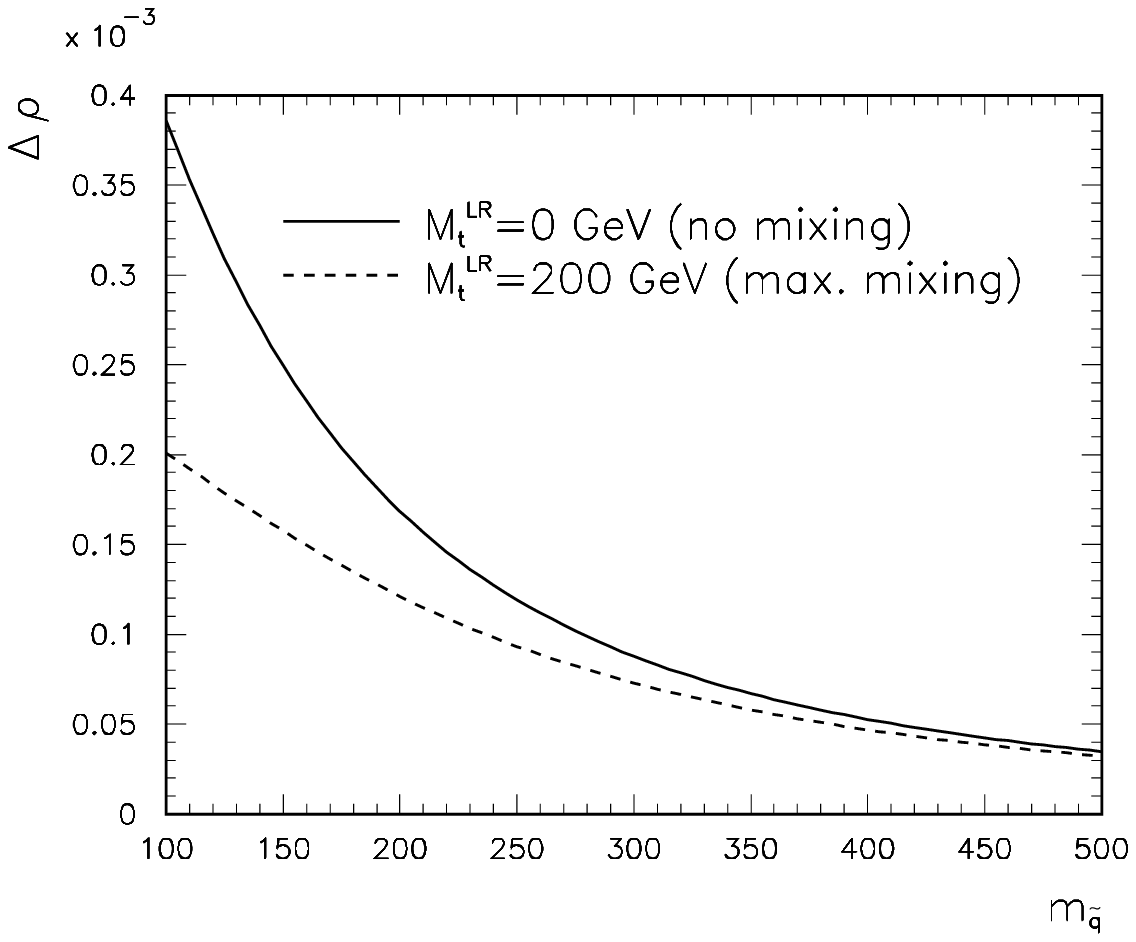,width=9cm,bbllx=120pt,bblly=270pt,%
                                        bburx=450pt,bbury=540pt}
     }
\parbox{16cm}{
\caption{\label{twoloopgluonplota}
$\dr^\SU_{\rm 1, gluon}$ as a function of $m_{\tilde q}$ for the scenarios
of Fig.~\ref{oneloopdrhoa}.
        }    }
\end{center}
\end{figure}

\begin{figure}[ht]
\begin{center}
\mbox{
\psfig{figure=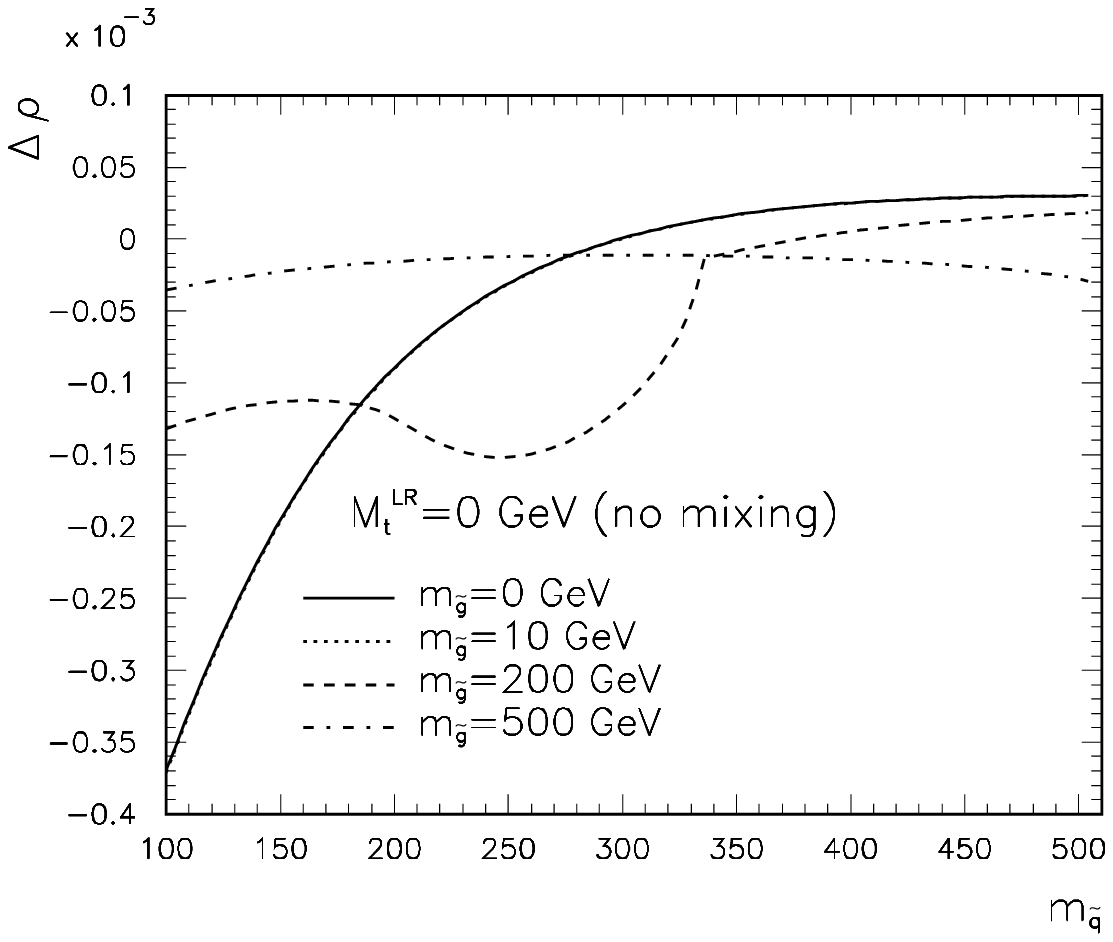,width=9cm,bbllx=120pt,bblly=270pt,%
                                         bburx=450pt,bbury=540pt}
     }
\parbox{16cm}{
\caption{\label{mglzeroplota}
$\dr_{\rm 1, gluino}^\SU$ as a function of $m_{\sq}$ in the no--mixing
scenario; $\Tb = 1.6$ and $\mgl = 0$, 10 (the plots are indistinguishable), 
200 and 500 GeV. 
        }    }
\end{center}
\end{figure}


\begin{figure}[ht]
\begin{center}
\mbox{
\psfig{figure=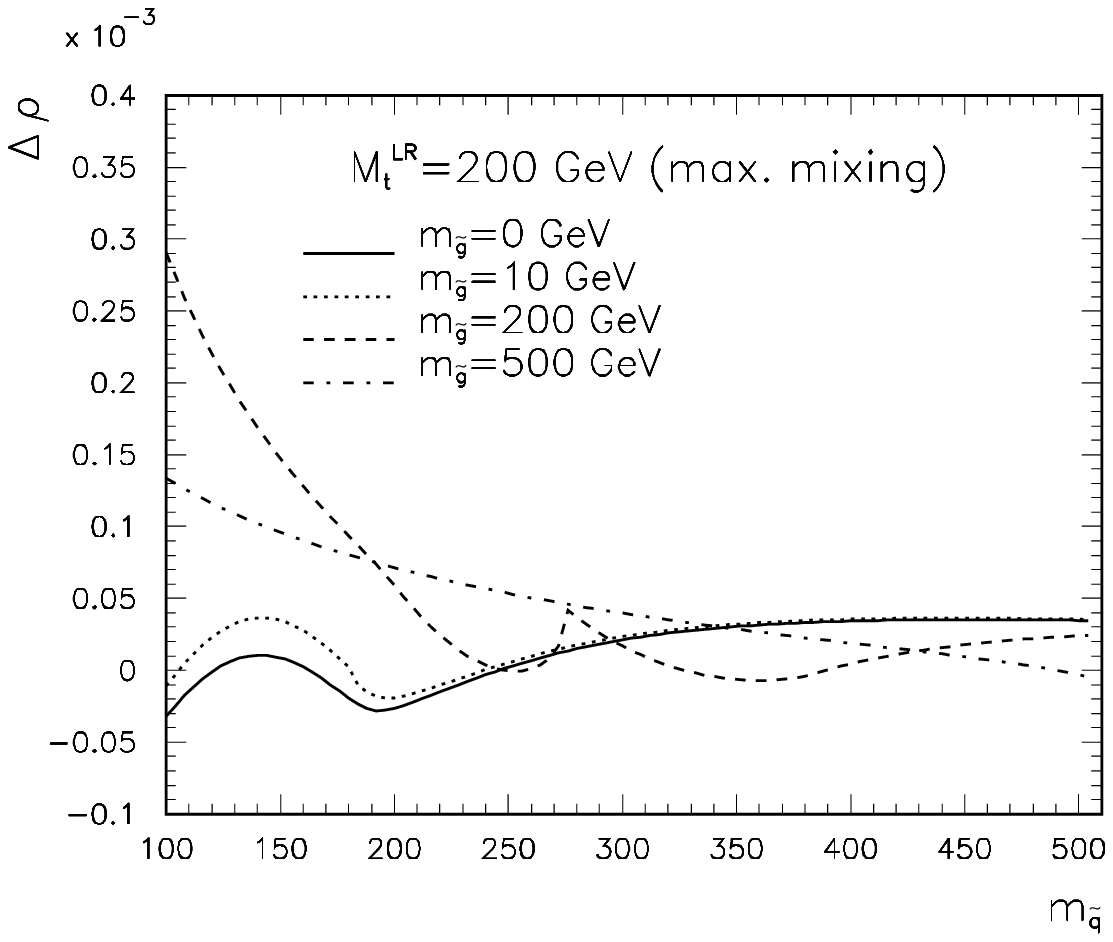,width=9cm,bbllx=120pt,bblly=270pt,%
                                         bburx=450pt,bbury=540pt}
     }
\parbox{15cm}{
\caption{\label{mglzeroplotb}
$\dr_{\rm 1, gluino}^\SU$ as a function of $m_{\sq}$ in the maximal
mixing scenario; $\Tb = 1.6$ and $\mgl = 0$, 10, 200, and 500 GeV.
        }    }
\end{center}
\end{figure}

\begin{figure}[ht]
\begin{center}
\mbox{
\psfig{figure=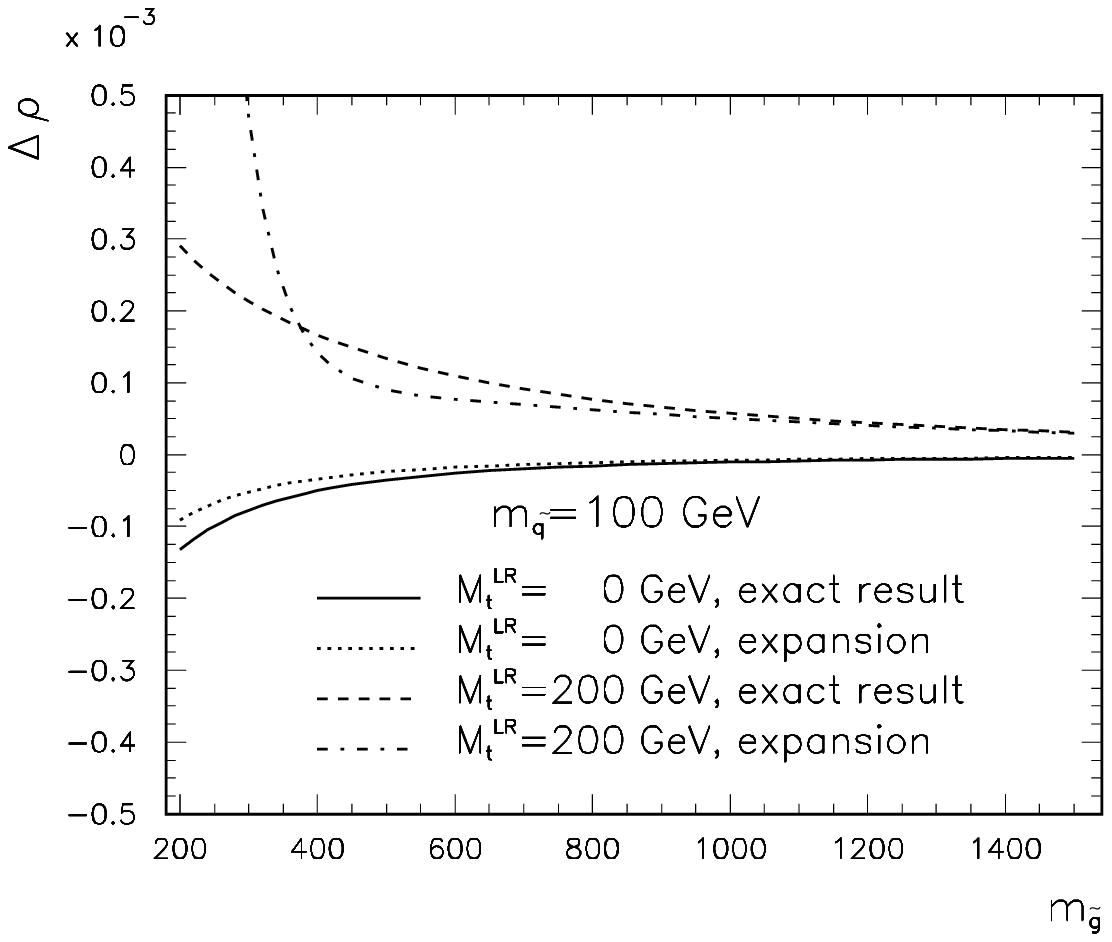,width=9cm,bbllx=120pt,bblly=270pt,%
                                         bburx=450pt,bbury=540pt}
     }
\parbox{16cm}{
\caption{\label{mglseriesplota}
Comparison between the exact result for 
$\dr_{\rm 1, gluino}^\SU$ and the result of the expansion 
eq.~(\ref{heavygluino}) up to ${\cal O}(1/\mgl^3)$ for 
the two scenarios of Fig.~\ref{oneloopdrhoa}; $m_{\sq} = 100$ GeV.
        }    }
\end{center}
\end{figure}


\begin{figure}[ht]
\begin{center}
\mbox{
\psfig{figure=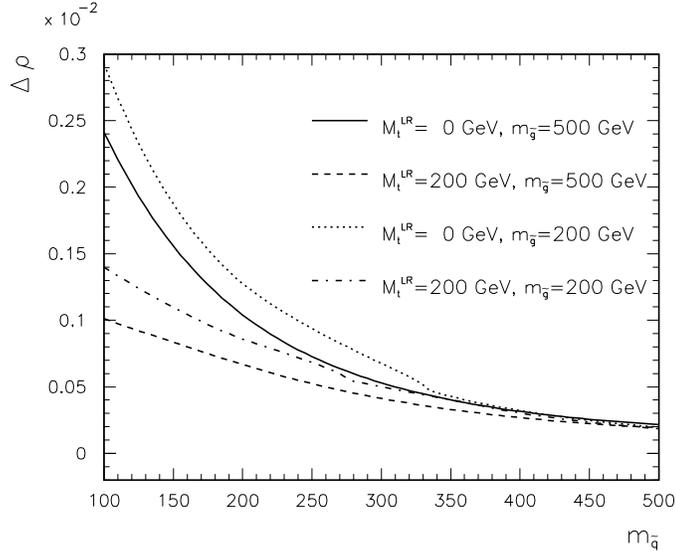,width=9cm,bbllx=120pt,bblly=270pt,%
                                        bburx=450pt,bbury=540pt}
     }
\parbox{16cm}{
\caption{\label{su2gut.ps}
$\Delta\rho_0^{\rm SUSY}$ calculated with one--loop corrected squark
masses as a function of $\msq$ 
for $\theta_{\tilde t} = 0$ and $\theta_{\tilde t} \sim -\pi/4$ ($\Tb
= 1.6$, $M^{LR}_t = 0 \mbox{ or } 200  
\mbox{ GeV}$) and two values of $\mgl$. 
        }    }
\end{center}
\end{figure}

\begin{figure}[ht]
\begin{center}
\mbox{
\psfig{figure=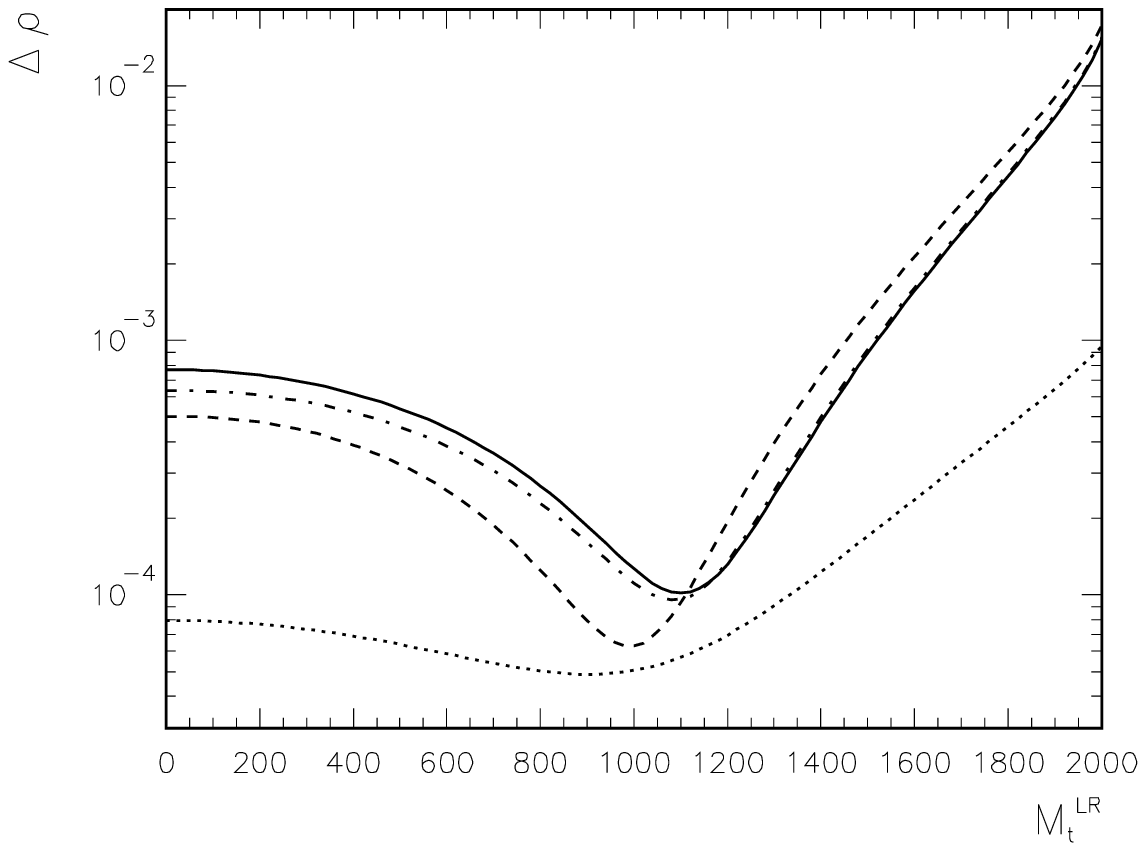,width=9cm,bbllx=120pt,bblly=270pt,%
                                        bburx=450pt,bbury=540pt}
     }
\parbox{16cm}{
\caption{\label{su2nogutlog.ps}
$\dr_0^\SU$ as a function of $M_{t}^{LR}$  for 
$\Tb=1.6$ and for $M_{\tilde{t}_L} / M_{\tilde{t}_R}=
1000/300$ (dotted line, the lines for tree--level and one--loop parameters 
are not distinguishable) and 300/1000 (solid line for tree--level 
parameters, for the one--loop parameters: dash--dotted for $\mgl=200$~GeV 
and dashed for $\mgl=500$ GeV. 
        }    }
\end{center}
\end{figure}

%% file: drholong.bbl
\newpage

\begin{thebibliography}{00}  

\bibitem{R1} For reviews on SUSY, see: H.P.~Nilles, Phys. Rep. 
{\bf 110}, 1 (1984); \\ 
H.E.~Haber and G.L.~Kane, Phys. Rep. {\bf 117}, 75 (1985); \\  
R.~Barbieri, Riv. Nuovo Cim. {\bf 11}, 1 (1988). 

\bibitem{R2} 
J.~Ellis, hep-ph/9611254; \\
S.~Dawson, hep-ph/9612229; \\
M.~Drees, hep-ph/9611409 ;\\
J.~Gunion, hep-ph/9704349 ; \\
H.E.~Haber, hep-ph/9306207. 

\bibitem{R3} Particle Data Group, Phys. Rev. {\bf D54}, 1 (1996).

\bibitem{R4} F. Abe et al., CDF Coll., Phys. Rev. Lett. {\bf 74} (1995) 2626;
\\ S. Abachi et al., D0 Coll., Phys. Rev. Lett. {\bf 74} (1995) 2632.

\bibitem{R5} S.~Bertolini et al., Nucl. Phys. {\bf B 353}, 591 (1991);\\
R.~Barbieri and G.~Giudice, Phys. Lett. {\bf B 309}, 86 (1993);\\
F. Borzumati, M. Olechowski and S. Pokorski, 
Phys. Lett. {\bf B 349} (1995) 311.

\bibitem{R6a} A.~Djouadi, G.~Girardi, W.~Hollik, F.~Renard and 
C.~Verzegnassi, Nucl. Phys. {\bf B 349} (1991) 48;\\
M.~Boulware and D.~Finnell, Phys. Rev. {\bf D44} (1991) 2054.

\bibitem{R6b} See e.g.\ W.~de Boer et al., hep-ph/9609209; \\
J. Wells, C. Kolda and G.L. Kane, Phys. Lett. {\bf 338} (1994) 219; \\
D.~Garcia, R.~Jim\'enez and J.~Sol\`a, Phys. Lett. {\bf B 347} (1995)
309; {\bf B 347} (1995) 321;\\
D.~Garcia and J.~Sol\`a, Phys. Lett. {\bf B 357} (1995) 349;\\
A.~Dabelstein, W.~Hollik and W.~M\"osle, in Perspectives for
Electroweak Interactions in $e^+e^-$ Collisions, ed.\ B.A.~Kniehl,
World Scientific 1995 (p. 345);\\
P.~Chankowski and S.~Pokorski, Nucl. Phys. {\bf B 475} (1996) 3.

\bibitem{R7a} R.~Barbieri and L. Maiani,  Nucl. Phys. {\bf B 224} 32 (1983); \\
C. S. Lim, T. Inami and N. Sakai, Phys. Rev. {\bf D} 29 (1984) 1488; \\
E. Eliasson, Phys. Lett. {\bf B 147} (1984) 65. ; \\
Z. Hioki, Prog. Theo. Phys. {\bf 73} (1985) 1283; \\
J. A. Grifols and J. Sola, Nucl. Phys. {\bf B 253} (1985) 47; \\
B. Lynn, M. Peskin and R. Stuart, CERN Report 86--02, p. 90; \\
R. Barbieri, M. Frigeni, F. Giuliani and H.E. Haber, Nucl. Phys. {\bf
B 341} (1990) 309; \\   
A. Bilal, J. Ellis and G. Fogli, Phys. Lett.  {\bf B246} (1990) 459; \\
M. Drees and K. Hagiwara, Phys. Rev. {\bf D 42}, 1709 (1990).

\bibitem{R7b} M.~Drees, K.~Hagiwara and A.~Yamada, Phys. Rev. {\bf D45},  
(1992) 1725; \\ 
P.~Chankowski, A.~Dabelstein, W.~Hollik, W.~M\"osle, S.~Pokorski and
J.~Rosiek, Nucl. Phys. {\bf B 417} (1994) 101;\\ 
D.~Garcia and J.~Sol\`a, Mod.\ Phys.\ Lett.\ {\bf A 9} (1994) 211.

\bibitem{hollik} W.~Hollik, Z.\ Phys.\ {\bf C 32} (1986) 291;\\
W.~Hollik, Z.\ Phys.\ {\bf C 37} (1988) 569.

\bibitem{R9} M.~Veltman, Nucl.\ Phys.\ {\bf B 123} (1977) 89. 

\bibitem{R10a} 
A.~Djouadi and C.~Verzegnassi, Phys.\ Lett.\ {\bf B195} (1987) 265;\\
A.~Djouadi, Nuovo Cim.\ {\bf A 100} (1988) 357.
 
\bibitem{R10b} 
B.A. Kniehl, J.H. K\"uhn and R.G. Stuart, Phys. Lett. {\bf B214} (1988) 621;\\
B.A. Kniehl, Nucl. Phys {\bf B347} (1990) 86;\\
A. Djouadi and P. Gambino, Phys. Rev. {\bf D 49} (1994) 3499; \\
L. Avdeev, J. Fleischer, S.M. Mikhailov and O. Tarasov, \\
Phys. Lett. {\bf B336} (1994) 560; E: Phys. Lett. {\bf B349} (1995) 597; \\
K.~Chetyrkin, J.~K\"uhn and M.~Steinhauser, Phys.\ Lett. {\bf B 351}
(1995) 331;\\
K.~Chetyrkin, J.~K\"uhn and M.~Steinhauser, Phys. Rev. Lett. {\bf 75}
(1995) 3394.

\bibitem{R10c}
Reports of the Working Group on Precision Calculations for the
Z-resonance, CERN Yellow Report, CERN 95-03, eds.\ D.\ Bardin, W.\
Hollik and G.\ Passarino.

\bibitem{R11} A.~Djouadi, P.~Gambino, S.~Heinemeyer, W.~Hollik,
C.~J\"unger and G.~Weiglein, Phys. Rev. Lett. {\bf 78} (1997) 3626.

\bibitem{R12} J.~Ellis and S.~Rudaz, Phys. Lett. {\bf B 128}, 248 (1983); \\
M. Drees and K. Hikasa, Phys. Lett. {\bf B 252}, 127 (1990).

\bibitem{R13a} G. 't Hooft and M. Veltman, Nucl. Phys. {\bf B 44} (1972) 189;\\
P. Breitenlohner and D. Maison, Commun. Math. Phys. {\bf 52} (1977) 11.

\bibitem{R13b} W. Siegel, Phys. Lett. {\bf B 84} (1979) 193; \\
D. M. Capper, D.R.T. Jones, P. van Nieuwenhuizen, Nucl. Phys. {\bf B
167} (1980) 479. 


\bibitem{PV} 
G.~'t~Hooft, M.~Veltman, Nucl. Phys. {\bf B 153} (1979) 365; \\
G.~Passarino, M.~Veltman, Nucl. Phys. {\bf B 160} (1979) 151.

\bibitem{B0eps} U.~Nierste, D.~M\"uller and M.~B\"ohm,
Z.\ Phys.\ {\bf C 57} (1993) 605. 

\bibitem{SUGRA}
V.~Barger, M.~Berger, P.~Ohmann, hep-ph/9409342; \\
W.~de~Boer, Prog. Part. Nucl. Phys. {\bf 33} (1994) 201.

\bibitem{PT} M.~Peskin and T.~Takeuchi, Phys.\ Rev.\ Lett.,
{\bf 65} (1990) 964; Phys.\ Rev.\ {\bf D 46} (1992) 381.

\bibitem{AB} G. Altarelli, R. Barbieri, Phys. Lett. {\bf B 253}, (1990) 161;\\
G. Altarelli, R. Barbieri, S. Jadach, Nucl. Phys. {\bf B 369} (1992) 3.

\bibitem{Schild} S.~Dittmaier, K.~Ko{\l}odziej, M.~Kuroda and
D.~Schildknecht, Nucl. Phys. {\bf B 426} (1994) 249, E: {\bf B 446}
(1995) 334;\\
S.~Dittmaier, M.~Kuroda and D.~Schildknecht,
Nucl. Phys. {\bf B 448} (1995) 3.

\bibitem{AB2} G.~Altarelli, 
talk at the XVIII Int.\ Symposium on Lepton--Photon Interactions,
Hamburg 1997, to appear in the proceedings.

\bibitem{EXP} J.~Timmermanns, talk at the XVIII Int.\ Symposium on
Lepton--Photon Interactions, Hamburg 1997, to appear in the
proceedings.


\bibitem{R13}
H.~Eberl, A.~Bartl and W.~Majerotto, Nucl. Phys. {\bf B 472} (1996) 481;\\
A.~Djouadi, W.~Hollik and C.~J\"unger, Phys. Rev. {\bf D 55} (1997)
6975;\\
W.~Beenakker, R.~H\"opker, T.~Plehn and  P.M.~Zerwas, Z. Phys. {\bf C
75} (1997) 349.

\bibitem{Davydychev} 
A.~Davydychev and J.B.~Tausk, Nucl.\ Phys.\ {\bf B 397} (1993) 123;\\
F.~Berends and J.B.~Tausk, Nucl.\ Phys.\ {\bf B 421} (1994) 456.

\bibitem{FA}
  J.~K\"ublbeck, M.~B\"ohm and A.~Denner, Comput. Phys. Commun {\bf
  60}, 165 (1990).

\bibitem{TC}
G.~Weiglein, R.~Scharf and M.~B\"ohm, Nucl. Phys. {\bf B 416} (1994) 606. 

\bibitem{ProcD} G.~Degrassi, S.~Fanchiotti and P.~Gambino, 
{\it ProcessDiagram}, a Mathematica package for two--loop integrals.

\bibitem{GK}
B.~Kniehl, Int. J. Mod. Phys. {\bf A 10} (1995) 443.

\bibitem{vienna}
A.~Bartl, H.~Eberl, K.~Hidaka, T.~Kon, W.~Majerotto, Y.~Yamada,
Phys.\ Lett.\ {\bf B 402} (1997) 303.

\end{thebibliography}
